# Risk Fluctuation Characteristics of Internet Finance: Combining Industry Characteristics with Ecological Value

Chuanmin Mi^[1], Nan Ye^[1], Runjie Xu^*[1], Tom Marshall[2], Yadong Xiao[1], and Hefan Shuai[3]

1.Nanjing University of Aeronautics and Astronautics, College of Economics and Management, Nanjing, 210016,China
2.Cambridge University, St Catharine's College, Cambridge, CB21RL, United Kingdom
3.Chinese University of Hong Kong, Department of Economics, Hong Kong, 999077, China

**ABSTRACT**

The Internet plays a key role in society and is vital to economic development. Due to the pressure of competition, most technology companies, including Internet finance companies, continue to explore new markets and new business. Funding subsidies and resource inputs have led to significant business income tendencies in financial statements. This tendency of business income is often manifested as part of the business loss or long-term unprofitability. We propose a risk change indicator (RFR) and compare the risk indicator of fourteen representative companies. This model combines extreme risk value with slope, and the combination method is simple and effective. The results of experiment show the potential of this model. The risk volatility of technology enterprises including Internet finance enterprises is highly cyclical, and the risk volatility of emerging Internet fintech companies is much higher than that of other technology companies.

---

**^Author Contributions: These three authors contributed equally to this work, all are the First Authors.**

*** Corresponding Author: Lilipota123@gmail.com (Runjie Xu).**

This research has been supported by the National Social Science Foundation of China (17BGL055) and E-commerce Innovation Space Fund (2019EC01). The author wants to thank Dr. Mahdi Miraz of the Centre for Financial Regulation and Economic Development of The Chinese University of Hong Kong for insightful comments and discussions. We are grateful to Dr. Qian-Yuan Tang (The University of Tokyo) for her help on the mathematical formula of our research. We acknowledge Runyu Meng at University of Science and Technology of China (USTC) for providing inspiring advice in researching method and paper writing. In terms of research ideas and the accumulation of previous research,this paper also benefited from the 2019 INFORMS international Conference on Service Science(ICSS) on 17-29 June,the 4th International Conference on Ambient Intelligence and Ergonomics in Asia(AmI&E) on 4-6 October, and the Machine Lawyering (https://www.legalanalytics.law.cuhk.edu.hk/).

# I. Introduction

The Internet is becoming an important infrastructure in society, providing technical support for people's communication, information search, data sharing, and so on (Warnick, B. (2018), Havrylchyk, O. et al (2017), Whaley (1993)). The Internet also brings opportunities and challenges for the business and financial industries, one of the most noticeable of which is Internet finance (Rau, P. R. (2018)).

Internet finance is an innovative form of finance that has grown faster than any other financial innovation in recent years. There is an essential difference between Internet finance and traditional finance. Internet finance is not a simple fusion between traditional financial institutions and the Internet, but is represented by Internet companies. Financial innovation based on Internet technology is more biased towards technology companies using information technology in order to improve the efficiency and model of financial services such as financing, investment and payment.

Internet finance is now a global phenomenon, and almost every country in the world has a relevant business model. We have listed representative Internet finance companies in major economic countries in Appendix 1. In many major economic systems, Internet financial platforms have replaced banks as the main means of capital circulation for consumers and SMEs.

Although Internet finance makes full use of the advantages of the Internet, it improves the efficiency of financial resource allocation to a certain extent and promotes the inclusive development of finance. However, Internet finance is still fundamentally financial in nature, and the use of information technology has strengthened the concealment, contagion, and contingency of Internet finance. The industry as a whole is basically in its infancy. Compared with the traditional regulatory system of traditional finance, many problems and hidden risks within the world of Internet finance have been exposed (Bruton et al (2015), Franks et al (2016)).

In the current environment, a large number of Internet companies compete with each other for user group traffic. This type of competition includes monopolizing the original business market and investing in new business markets. This has caused some businesses to be unprofitable or to put themselves at risk of future unprofitability. As a result, the financial industry has a greater risk exposure, and the risk value of Internet companies is much higher than that of traditional industries. This competitive business model is particularly evident in the Internet finance industry, and the role of Internet finance in the entire economy is becoming increasingly important (Meeker (2018), Segal (2016)).

The entire Internet industry adopts this unique method to reinforce its strength in continuous financing and acquisitions. This business model believes that value is created not only by producers, but also by customers and other members of their value creation ecosystem. From this point of view, a company in the Internet industry only needs to make strategic investments and acquisitions in a range of fields to obtain business and users in this field, and to use these businesses to deploy infrastructure to serve original users, thus strengthening its position in the digital economy.

However, these investments and acquisitions do not take profit as the primary purpose, but rather depend on whether the field is related to the current main business or whether it provides infrastructure, technology, services or products for its business development. This expands the business, facilities and workforce but, at the same time, it also brings huge risk exposure, and involves more other unpredictable cost risks, such as potential labour disputes, compliance costs, unprofitability, etc. On the other hand, existing unprofitable businesses may also be part of the Internet ecosystem, or they may be laying out strategies for future ecological structures. This business model concept may challenge the assumptions of traditional value creation and value acquisition theories (Barney (1991), Peteraf (1993)). According to the Internet industry, value creation is provided by both the supply side and the demand side. Value is created not only by producers but also by customers and other members of their value creation ecosystem.

From this perspective, competitive positions can be obtained from multiple perspectives, such as based on resource supply and user activity. This concept has led to frequent acquisitions and investment activities among the Internet industry.

Therefore, in studying the process of risk formation, we also need to consider the possible ecological value behind the risk. For the Internet industry, in addition to value estimates, there are also differences in cash flow, sales growth, R & D, and high-risk early warnings (Bartov et al (2002)). Internet finance is composed of both Internet technology and financial technology. It needs to comprehensively evaluate Internet risks based on corporate value, and it also needs to evaluate financial risks based on the Internet ecological foundation. The original risk research can no longer describe the Internet industry risks, especially in the field of Internet finance. This makes the entire study a very complex issue.

At present, there are few papers on the model, risks, or characteristics of Internet finance in major economic systems. The purpose of this article is to study risk fluctuations based on the consideration of corporate value disturbances.

First, we explain the tendency of Internet companies' business income (Internet finance also belongs to the Internet industry). This tendency comes from a mode of competition in the Internet industry: in simple terms, Internet companies do not consider whether their business is profitable and continue to increase new investment acquisitions. Its purpose is to achieve Internet user group competition, infrastructure construction and data collection.

We take into account that some Internet companies have excellent risk control capabilities, which makes them able to withstand the fluctuation of risks in a higher position while building their position in the digital economy. At this time, if the angle of risk mass is used to measure risk, distortion will often occur.

To address this challenge and mitigate the impact of diversified factors on risk, our goal is to propose an empirical method to measure the risk level of Internet companies – by generating a risk index to observe the risk level of individual companies and the risk level of the entire macro market. The total incremental risk of different companies at the same time determines the size of our industry risk index. The company with the highest contribution in the risk index is therefore the company with the highest risk. Conceptually, this calculation is similar to the stress test routinely applied to financial companies, but here it uses only publicly available information and is fast and cheap.

We start by building a database of all the acquisitions made by fourteen representative technology companies in their financial statements between 2015 and 2019.

Combined with the content proposed by Whaley (1993), we assume that market fluctuations can reflect investors' emotions and expectations. Negative or positive corporate news will spread to the market through a variety of channels, change investors' expectations of the company, and have a surge or fall on the current stock prices. We specify the level of enterprise risk volatility in a certain period of time as the basis for risk measurement.

To measure risk, we were inspired by the POT model to construct a generalized Pareto distribution to measure tail risk. The risk fluctuation range (RFR) is one of our findings. It eliminates the interference of the risk volume on the risk assessment results. This method focuses on the fluctuation of risk in time series to study the risk problem. We believe that for Internet companies, the level of risk means different strategic layouts and different future value expectations. Although the current acquisition of enterprises and businesses and other business models cannot be profitable and cause certain risks, they can provide infrastructure for the future ecological environment. Our method is compared with the original risk model, and we find that the original model only depicts the risk volume and cannot better reflect the characteristics of the enterprise on risk fluctuations.

In order to solve these potential problems, we have formed several additional comparative tests, including comparisons between Internet financial companies and other Internet companies, comparisons between Internet technology companies involving various businesses, and comprehensive index and corporate stock prices.

We pay attention to the local correlation of risks, realize the process from scale transformation to feature extraction, and obtain the features with overlapping risk cycles from it. We further found that the amount of risk does not accurately reflect the risk of Internet technology companies. Specifically, the high risk volume can indeed reflect that the enterprise itself is in crisis. However, while some enterprises have high risk indicators, their life span, market value, and user base are still in the leading position of Internet technology companies. This is the aforementioned relationship between ecological value and risk.

Our results show that companies with large risk values do not necessarily have the largest risk fluctuations. The value of the risk can only reflect the volume of the risk, but the risk fluctuation can show changes in investor confidence within a certain time frame, so it can better target Internet companies. Through this channel, Internet technology companies can learn about their own risk position in the industry, and regulatory agencies can observe the industry's overall risk dynamics in a timely manner in order to prevent and deal with the corresponding problems.

From our correlation test results, we can find that the correlation of risks is easily disturbed by market factors. Specifically, companies in the US market have a certain degree of similarity in risk fluctuations, and companies in the Chinese market have types of similarity. In summary, our results emphasize that the partial periodic overlap of risk volatility is an important feature that is difficult to decouple within the Internet industry.

We believe that the risk exposure of Internet technology companies is caused by a variety of factors. This factor not only includes multiple influences such as business models, market environment, and macroeconomics, but also is affected by corporate strategy. We have further proved that our results are effective for the risk check of Internet technology companies through comparative experiments. The results of this research can be used for risk analysis in the same type of industry, which can have a good overview of the cyclical overlap of risks between industries, and thus have a clearer understanding of the overall risk of the entire industry.

# II. Main Forms of Internet Finance

The main forms of Internet finance include Internet payment, Internet lending, Internet crowdfunding, Internet fund, Internet insurance, and Internet trust (Hou (2016), Gathergood (2017), Jagtiani (2017), Hong (2017), Xu (2020)).

In the process of real market operation, there exists the situation that a single enterprise focuses on a certain format, and there is also the situation that an organization is involved in multiple formats. There is the penetration of Internet enterprises into financial business, and there is also the use of Internet technology by traditional financial institutions to compete in the market. There are formal financial enterprises with licenses, but there are also "Shadow Banks" free from the private sector.

## *A. Internet Finance for Payment*

Internet payment, known as mobile payment, generally refers to the service of initiating and executing the transfer and use of funds through terminals (mobile phones, PCS, etc.).

The development of Internet payment can be summarized as follows:

Firstly, Internet financial institutions use the marketing strategy of two-way subsidies to merchants and consumers to promote offline merchants to open mobile payment functions.

Secondly, Internet financial institutions have opened the multi-country currency payment function and started the mobile payment function in domestic and overseas consumer markets.

Thirdly, Internet financial institutions and personal credit investigation are linked to build a credit consumption system, using artificial intelligence and big data to gradually improve the online payment environment based on the credit system, providing a technical analysis basis for microfinance and credit consumption.

## *B. Internet Finance for Lending*

Internet lending is generally embodied in two forms:

One is P2P lending, an online platform for financial services initiated by Internet companies. The platform provides matching services for users with financing or investment needs.

The other is the internetization of traditional microfinance, which is initiated by traditional financial institutions and extends the source of customers through the assistance of Internet technology, which is an improvement in efficiency.

However, Internet lending also poses certain risks to the Internet financial environment:

(a) From the perspective of the business model of Internet lending, its innovation lies in the use of point-to-point matchmaking, where investors remit funds to the platform and lend them to the demander through the platform. Such a business model also brings capital risk easily. Firstly, users' funds are deposited in the online loan platform, and the credibility of the online loan platform is crucial. Secondly, if the borrower defaults or breaks his promise, it will have an impact on the online loan platform itself. Thirdly, stock market volatility will often cause the withdrawal effect of the online lending platform. If the platform's capital strength and risk control ability are weak, negative news will easily cause a run.

(b) The main users of Internet lending services are small and micro enterprises or individual consumers, and the users are often from the objects not served by traditional financial institutions. Internet companies attach great importance to the user base, compete in the market through subsidies, and have low barriers to entry for customers; this means the customer quality of Internet lending carries risk attributes.

(c) In recent years, Internet lending has attracted much attention from the market, and the surge of relevant operating platforms has further intensified the survival pressure of small and medium-sized platforms in the main competitive markets, leading to poor operation, suspension of business and even fraud of some platforms.

(d) In the process of competing in the Internet lending market, native Internet companies

operate without licenses, which is essentially regulatory arbitrage. Some enterprises have a weak awareness of risk control, and their technical conditions fail to meet requirements. After the gradual tightening of the regulatory layer at the later stage, they have accelerated the preferential selection in the market. As a result, some platforms were closed down, the capital chain was broken, and the interests of customers were damaged.

## *C. Internet Finance for Crowdfunding*

Internet crowdfunding generally refers to the use of an Internet platform for open micro-financing activities by a project, company or financier (while paying for equity, products or services). Compared with traditional financing methods, crowdfunding is more open, and the availability of funds is no longer based on the commercial value of the project. Any project that users are interested in can be funded through crowdfunding.

Crowdfunding platforms like AngelList, Fundable, Crowdfunder, and EquityNet can also increase the rate of information sharing, negotiation, and fundraising.

## *D. Internet Finance for Fund*

Internet fund refers to the direct communication between investment clients and third-party financial institutions on the basis of Internet media, so as to bypass the involvement of banks, which is an extension and supplement of traditional financial services. Under this "financial disintermediation" financial management mode, banks no longer play a paid connection role between customers and third-party financial management institutions, weakening the financial intermediary status of banks, greatly improving financial management efficiency and reducing financial management cost.

Ant Financial's Yu’E Bao was the earliest Internet fund. The combination of fund and Internet not only changes the sales mode, but also ensures Internet fund financing attains the same characteristics of high liquidity, high security and high profitability as traditional fund financing, whilst gaining additional characteristics different from traditional fund financing mode.

Firstly, Internet funds rely on modern information technologies such as big data, social networks and mobile payments to virtualize trading venues. Compared with the traditional fund financing mode, they facilitate the funds' business operations, reduce operating costs, and greatly improve business efficiency.

Secondly, the efficient matching of fund products and customer investment needs can be realized. Under the Internet fund financing mode, investors can master more information conducive to their own investment through the network platform and can easily complete the comparison of various fund products, so as to screen out suitable high-quality investment targets. A more important part of the Internet finance model is that Internet fund financing enables the vast majority of people (especially those from low-income groups) to participate in such financial innovation activities, effectively alleviating financial exclusion and justifying the label of inclusive finance.

## *E. Internet Finance for Insurance*

Internet insurance is an emerging insurance marketing model based on computer Internet, which is different from the traditional insurance agent marketing model.

Compared with the traditional way of insurance promotion, Internet insurance allows customers to choose their own products. Customers can compare the products of multiple insurance companies online, with transparent premiums and clear protection of rights and interests, which can greatly reduce the surrender rate of traditional insurance sales. The network can promote the accelerated development of the traditional insurance industry, so that the selection of insurance, insurance plan design and marketing costs reduce, which is conducive to improve the operating efficiency of insurance companies.

## *F. Internet Finance for Trust*

Internet trust is a new model of hot Internet finance in recent years. It is the combination of a P2B (person to business) financial industry investment and financing model and O2O (offline to online) e-commerce model to verify the offline information of borrowing enterprises, pledge assets, credit rating and other credit investigation services, so as to ensure the financial security of investors.

Compared with traditional trust business, Internet trust has the following advantages: firstly, the use of Internet technology to disclose information on trust products in a timely fashion in order to improve the transparency of the operation of trust products and better fit the needs of users; secondly, drawing from the real needs of customers, the improvements in service through the Internet platform; and thirdly, the use of Internet technology to provide quality all-round services for the stock of customers, such as net value management and coupon reminder.

## *G. Internet Finance for Consumption*

This refers to the provision of financial services with information, electronic and data characteristics for each link or stage of consumption through Internet technology and platform, including but not limited to payment, financing, liquidation, etc.

The Internet consumer finance model is divided into vertical installment purchase platform and e-commerce consumer finance platform.

### *G.1. Vertical installment purchase platform*

Installment purchase platforms cooperate with online and offline merchants to integrate loan application and instalment into the consumption process. When consumers need to purchase goods and services from merchants, they can apply for instalment credit from instalment purchase platforms. After the application is approved, merchants can immediately provide corresponding products or services. This method mainly cuts into the consumption scene with low traditional consumer finance penetration through installment service or consumer loan, but its defect lies in high risk control requirements. Since it is difficult to have the advantage of e-commerce big data for a certain vertical market segment, most installment purchase platforms rely on Internet credit investigation and have a high bad debt rate.

### *G.2. E-commerce consumer finance platform*

By providing customers with installment payment, they can consume on the platform, mainly represented by the Tokio of Ant Financial and the baitiao of Jingdong. After consumers consume commodities on the e-commerce platform, the platform shall provide funds to pay, and the platform merchants shall issue commodities to consumers, after which the consumers shall repay the loans. Because it can grasp the customer's purchase records, capital flow and logistics information, the model can accurately obtain customer portraits, thereby reducing the cost of risk control. With the expansion of e-commerce consumer finance platforms, financial business is gradually separated from the main business, forming an independent plate.

# III. Overview of Internet financial risks

Measuring, predicting and controlling risk is the core of financial economics theory and practice (Whaley (1993)). In previous studies, we constructed a complex network of Internet finance and explained the correlation between risks. This section will introduce the risks of Internet finance based on previous studies.

Based on the research literature on financial risks and Internet financial risks, Internet financial risks mainly include technical risks, operational risks, legal risks, credit risks and business risks (Xu (2020)).

## A. Technical Risk

Technical risks include operating system vulnerabilities, Trojan viruses, internal information leakage, identity forgery login, network transmission problems, server maintenance, and natural disaster damage. Similar to the risks of traditional finance, all risks in Internet finance are contagious (Amini (2016)). For example, operating system vulnerabilities will lead to Trojan virus waiting for an opportunity to invade the server resulting in abnormal and internal information leakage, network transmission problems will lead to Trojan attacks and internal information leakage, improper maintenance of the server will lead to network transmission problems. From the perspective of external impact, technical risk types will produce cross-class contagion with other risk types, mainly reflected in the impact on operational risk types and legal risk types. For example, operating system vulnerabilities, network transmission problems, etc., can lead to malicious intrusion by others (operational risk type). Internal information leakage may lead to the abuse of personal information, resulting in user lawsuits.

## B. Operational Risks

Operational risk mainly exists in the business model of Internet finance, including internal operational risk, malicious intrusion risk, user accidental operation risk, service provider operation risk, outsourcing technology risk and technology cooperation development risk. Some subdivided risk factors can infect each other. For example, the relationship deteriorates during the process of service provider, outsourcing technology or technical cooperation development, which may lead to the risk of malicious intrusion by partners. From the perspective of external impact, operational risk types can cause cross-class contagion with other risk types, mainly reflected in technical risk types, legal risk types, business risk types and enterprise operation risk types. For example, when Internet financial enterprises have disputes with service providers and outsourcing partners, due to the completeness of existing laws, they cannot be effectively investigated for responsibility, which may affect the business development, reduce the trust of users and put pressure on the operation of the whole enterprise.

## *C. Legal Risks*

Legal risks include risk of incomplete information disclosure, risk of abuse of personal information, risk of illegal financing, risk of complete legal protection, risk of illegal business operation, risk of user prosecution and risk of national policy. There can be contagion between risk segments. For example, the perfection of laws and regulations makes it impossible for Internet financial enterprises to cover up their abuse of personal information. After the event is exposed, they may also suffer from public relations crisis and user lawsuits. From the perspective of external impact, legal risk types can cause cross-class contagion with other risk types, mainly reflected in the types of credit risk, business risk and enterprise operation risk. For example, compared with traditional finance, the operation of Internet finance suffers from lack of industry standards; its ability to resist risks such as capital flow, market cycle and interest rate is very low; and the platform is prone to credit violation risks.

## *D. Credit Risk*

Credit risk is ubiquitous in Internet finance and an important part of risk prevention (Smedlund (2012), Leduc (2017)). This type includes term mismatch risk, contract default risk and false publicity risk. Subdivided risk factors can infect each other. For example, the investment asset maturity of Internet financial products is long, while the liability maturity is short, so that the financial products of Internet financial companies may not be able to pay in time, resulting in the risk of maturity mismatch which will eventually evolve into credit default. From the perspective of external impact, credit risk can cause cross-class contagion with other risks, mainly reflected in the types of legal risk, business risk and enterprise operation risk. For example, after the risk of contract default of Internet finance platforms, there may be lawsuits or punishments from regulators, which will reduce the confidence of Internet finance business in users' minds and thus affect the development of business activities.

## *E. Business Risks*

Business risks include capital flow risk, market cycle risk, interest rate risk, user preference risk and investor relationship risk. There will be some degree of contagion between the segmented risks. For example, Internet financial enterprises cannot obtain sufficient funds at reasonable cost in time to cope with the growth of assets or to pay the debts due, thus causing the risk of capital chain breaking, which may be caused by investor relationships, as well as interest rate risk and market cycle. From the perspective of external impact, business risk types will generate cross-class contagion with other risk types, mainly reflected in credit risk types and enterprise operation risk types. For example, the performance of Internet finance affects the health of its capital flow. This is because a large number of platforms attract users through financing and subsidies in the process of business development, which makes the health of capital flow within the platform extremely important.

## *F. Risks to Traditional Finance*

Internet finance can realize the network matching between the two sides of funds, without any financial intermediary. This has a certain impact on the business of traditional finance, and Internet finance’s rapidity and low threshold makes it an attractive replacement possibility for the traditional financial medium.

The payment and settlement functions of Internet finance belong to the basic services of banks and contribute important non-interest income to banks. The new payment and settlement channels provided by Internet finance have affected the intermediary business of traditional finance.

Internet finance research and development of information processing under the big data technology provide better risk management. Information was originally the core element in the process of financial resources allocation. The information processing methods of traditional finance were mainly made mandatory by third-party professional rating agencies and regulatory authorities under the guidance of the government, so the cost of information acquisition was extremely high. Internet finance has a wide range of data sources. Through in-depth data mining, Internet finance can improve risk control capability at a lower cost (Haldane (2011), Georg (2013)).

There is a risk generation mechanism between Internet finance and traditional finance. Through the impact of Internet lending business, Internet fund business and Internet payment business on traditional banks, they have successfully squeezed the assets, liabilities and intermediate business of commercial banks, resulting in higher operating costs, lower profitability, loss of deposits and higher leverage ratio. The services provided by Internet finance to users have changed the level of money supply and demand and interest rate and have had an impact on the intermediary variables of the macro economy. This led to the expansion of bank credit, the decline in the demand for money, the weakening of the effectiveness of macro-control, and finally to the banking crisis.

There is a risk contagion mechanism between Internet finance and traditional finance. Internet finance uses the channels of funds generated by cooperation with commercial banking businesses, and through various platforms of the Internet finance industry, based on the combination of bank electronic accounts, deposit accounts, depository accounts, and reserve accounts, and through other methods, the Internet finance as a whole is integrated. Industry risks, service target risks, legal regulatory risks and technical operation risks are transmitted to commercial banks. Internet financial users continue to grow in number and to become gradually dependent on the platform. If the payment service provided by the Internet platform is suddenly paralyzed and not repaired in time, the "herd effect" caused by crisis news via media channels may trigger investors to change their investment expectations, run on cash and other social situations (Glasserman (2015)).

# IV. M & A and Investment in Internet Finance

Since the Internet boomed in the mid-1990s, Internet companies have been trying new business models to achieve their goals. The behaviour of mergers, acquisitions and investment as a kind of business model directly affects the entire Internet industry’s continuation and use of this unique way to promote the development of the entire industry. Internet financial companies also inherited this characteristic, namely, strengthening their own power in continuous financing and acquisitions.

In the Internet industry, the exploration of business models directly determines competitiveness and becomes a strategic focus for managers in different industries. In recent years, the Internet industry believes that a model different from traditional industries can stimulate user interest and may become a source of excess returns. Rumors of exceptional profitability from innovative business models are not uncommon. Take Google as an example: the company went to prosperity with a paid listing advertising business model. Xerox, meanwhile, chose to lease its copy machine instead of selling it, making the company one of the most profitable companies at the time (Chesbrough (2007), IBM (2006), Ireland (2001), Johnson (2008)).

As Internet technology blurs the differences between industries, lowers the barriers to entry and leads to more intense competition, Internet companies are forced to choose to directly obtain the right to use a technology or a business through investment and acquisition. This trend has created an ecological environment for investment and mergers in the existing Internet industry.

We believe that while the investment and acquisition of the Internet industry increases the individual's competitive position in the industry, it will also affect the uncertainty risk of the entire industry. The risk of a company is rapidly spread throughout the industry, so this risk will be reflected in the trend similarity in risk volatility.

We use the financial statements of Internet companies to explore the relationship between investment M & A models and risks. As shown in Figure 1 below, we have selected Alibaba as the reference object (the other thirteen Internet companies are Ebay, Facebook, Paypal, Google, Apple, Twitter, Amazon, BIDU, JD, Tencent, YRD, PPDF, and DNJR, detailed further in the "Data" section).

In Alibaba's financial report analysis, we can clearly find that research, development, sales and other expenses have steadily increased, while interest and investment income and income from operations have declined. This seems to indicate that the company's investment cannot be translated into actual returns. However, the company's revenue has risen at a faster rate, which indicates that the overall revenue and expenditure situation is more optimistic.

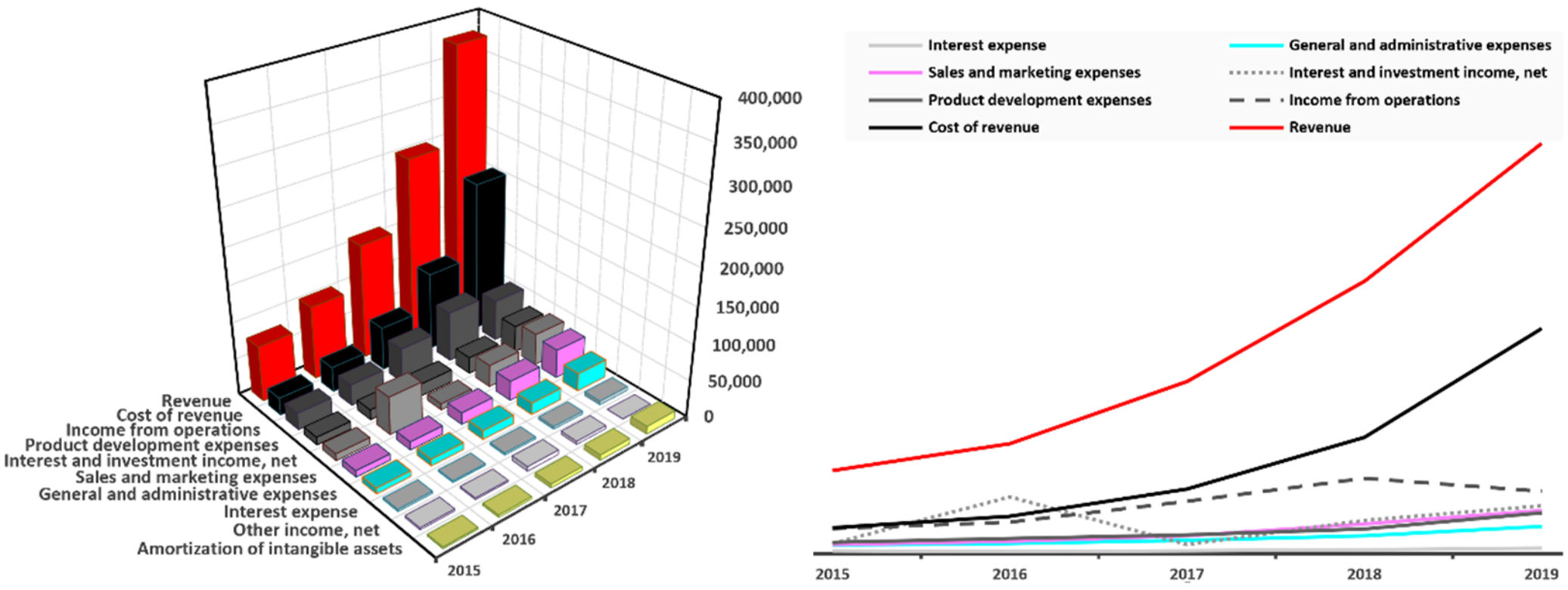

**Figure 1. Data from Alibaba's operating statements for the 2015-2019 financial years**

Alibaba, as China's largest e-commerce company, is also a representative Internet company in China. This type of company attaches great importance to the long-term benefits of the digital economy, as well as the acquisition of network traffic, so it continues to increase expenditures on business, investment acquisitions and strategy.

Judging from the financial reports in recent years, many of the company's newly invested businesses have low or negative profit margins, and the invested businesses are often in the early stages of exploration. Many of these business models are more efficient in attracting and converting paying merchants low. These investments and acquisitions did not increase Alibaba's revenue. Instead, from FY 2018 to FY 2019, Alibaba's adjusted EBITDA profit margin fell from 42% to 32%. Table 1 shows the companies acquired or invested by Alibaba and the related businesses they opened. The types of these investments and initiatives include:

(a) Commercial products capable of expanding and strengthening core competitiveness, including supporting Alibaba's logistics network, local service business, new retail plans, direct sales, and cross-border e-commerce

(b) Expanding construction of various facilities and increasing the number of employees.

(c) Researching and developing new technologies to improve technological infrastructure and cloud computing capabilities;

(d) Innovative measures for digital media and entertainment business.

That is, according to traditional theories of strategy, such as the resource-based view of the firm or the positioning view, value creation is a supply-side phenomenon in which value is created exclusively by producers, not by customers; and competitive advantage is single-sourced resource-based only or activities-based only (Massa (2017)).

The concept of business model in the Internet industry may challenge the assumptions of traditional value creation and value acquisition theories. In contrast with the concept based on the Internet business model, value creation is provided by both the supply side and the demand side. Value is created not only by producers but also by customers and other members of their value-creating ecosystem. From this perspective, competitive position can be obtained from multiple perspectives, such as resource availability and user activity.

Therefore, we see Alibaba's strategic investment and acquisitions in a range of areas as strengthening its digital economy leadership in China. These investments and acquisitions do not take profit as the primary purpose, but rather depend on whether the field is related to the current main business or whether it provides infrastructure, technology, services or products for its business development. These products can promote user activities and continue to create value for the entire ecosystem.

However, such strategic investments and acquisitions still have an adverse impact on Alibaba's financial performance. For example, if you acquire a company with a lower profit margin or a loss, such as the acquisition of a controlling stake in a company such as Lazada and Cainiao Network, these acquired loss companies may not be profitable at all in the future. In addition, Alibaba's investment to expand its business, facilities and workforce will also involve more other unpredictable cost risks, such as potential labour disputes, compliance costs and risks.

In the process of considering risks, we must consider the possible ecological value behind the risks. Therefore, how to comprehensively judge the relationship between the value and risk of Internet companies is a very important research question.

## Table I

## Main Business of Alibaba

| Core Commerce | Key Businesses | Key Services |
|---|---|---|
| Retail Commerce | Taobao Marketplace, Tmall, Freshippo, Intime Department Store, AliHealth, Lazada, AliExpress, Tmall World, Trendyol, Daraz | Innovative Supply Chain, Intelligent Medicine and Internet Healthcare, Seamless Shopping Experience, Digitalized Operating Systems, In-Store Technology, Supply Chain Systems, Consumer Insights and Mobile Ecosystem |
| Wholesale Commerce | LST, 1688, Alibaba | Digital Sourcing Platform for Retailers |
| Logistics Services | Cainiao Network | Logistics Data Platform, Global Fulfillment Network, One-Stop-Shop Logistics Services and Supply Chain Management Solutions |
| Consumer Services | Ele.me, Koubei, Fliggy | Online Travel Platform, Ordering Food and Beverages |
| Cloud Computing | Alibaba Cloud, UC Browser, Alimama | Elastic Computing, Database, Storage, Network Virtualization Services, Large Scale Computing, Security, Management and Application Services, Big Data Analytics, Machine Learning Platform, and IoT Services |
| Digital Media | YouKu, UC Browser, Alibaba Pictures, Damai, Alibaba Music, Alibaba Literature | News Feeds, Literature, Music, Online Video Platform, Mobile Browsers, Cinema Ticketing Management and Data Services for Entertainment Industry |
| Innovation Initiatives | Amap, DingTalk, TmallGenie | Navigation, Food Delivery and Taxi-Hailing Services |
| Financial Technology | Ant Financial | Digital Payment, Wealth Management, Insurance and Credit |

# V. DATA

These risks that will occur, or have already occurred, will be transmitted to the market through other means, such as negative news, and cause investors to change their investment expectations for the enterprise. Ultimately, the risks will be reflected in changes in stock prices. We selected stock price data of technology companies to build a risk model.

According to different main businesses and the degree of attention given by the market, we selected fourteen Chinese and American technology companies: Alibaba, JD, Facebook, PayPal, Ebay, Google, Apple, Twitter, Amazon, BIDU, Tencent, YRD, PPDF, and DNJR.

Some Internet finance companies, despite their high market share and influence, have not gone public. These companies often come from large tech companies. For example, Ant Financial is from Alibaba, WeChat Pay is from Tencent, Jingdong Baitiao is from Jingdong, and so on. We chose to include their parent company in our comparison. It is worth noting that due to China's regulatory issues, it is difficult for Chinese technology companies to list in China, so most technology companies have chosen the ADR method to list in the United States.

Among the fourteen technology companies, we listed YRD, PPDF, and DNJR, three Internet finance companies whose main business is lending, and compared them horizontally with other companies. The reason is that the predecessors of these Internet financial companies did not come from a large technology company. It is worth noting that the Internet financial company DNJR went public on March 20, 2018; in recent months, due to the suspected overdue payment, it was reviewed by the regulatory authorities and was suspended by Nasdaq on November 4, 2019.

In addition, we conducted risk volume comparison experiments to find that the market in which the main business is located will have an impact on the risk trend. For example, Alibaba and JD in China and Facebook and Paypal in the United States have almost similar trends in risk volume. Therefore, we will consider the impact of the overall market environment and macroeconomic development on the judgment of risk trends. According to the securities market, where the company is listed and the important securities markets of the main business countries, we have added S&P 500 (Standard & Poor's 500), DJIA (Dow Jones Industrial Average), National Association of Securities Dealers Automated Quotations (Nasdaq), Shanghai Securities Composite Index (SSE), Hang Seng Index (HIS), and Shenzhen Securities Component Index (SZI).

In addition, before building the risk analysis model, we compiled the acquisition activities of fourteen companies between 2015 and 2019. All the information below comes from the company's financial statements. See “Appendix 1” for details.

# VI. Risk volatility model (RFR)

In the introduction it was mentioned that Internet finance is composed of Internet technology and financial technology. It needs to comprehensively evaluate risks based on the value of the enterprise and also to evaluate risks based on the Internet ecological foundation. This is because the value of the Internet industry is created not only by producers, but also by customers and other members of their value creation ecosystem. Therefore, in studying the process of risk formation, we also need to consider the possible ecological value behind the risk.

Therefore, the original risk research has not been able to describe the risks of the Internet industry, especially the risks in the field of Internet finance. We take into account that some Internet companies have excellent risk control capabilities, which makes them able to withstand the fluctuation of risks in a higher position while building their position in the digital economy. At this time, if the angle of risk mass is used to measure risk, distortion will often occur. In order to eliminate the interference of the risk volume on the risk assessment results, we focus on the risk fluctuations in the time series to study the risk problem.

Although there is a large amount of literature to study risk fluctuations, such as ARCH and GARCH models in traditional finance (Engle et al (1982), Bollerslev  et al(1986)), DFA methods to detect the autocorrelation of time series, and theoretical models to explore the micro mechanism of "volatility aggregation" phenomenon, etc (Xu (2005)). Most theories believe that price volatility can represent the financial market's response to external stimuli (such as emergencies or abnormal information) and is one of the important indicators of financial risk. (Feng et al (2012))

This article differs from other risk fluctuations in that our risk fluctuations focus more on the trajectory of corporate risk on a certain level. We call it the risk volatility (RFR).

We believe that for Internet companies, the level of risk means different strategic layouts and different future value expectations. Although the currently acquired companies and businesses and other business models are not profitable and cause certain risks, they can provide a future ecological environment and provide infrastructure construction. Therefore, only by excluding the interference of the risk volume, can a comparative analysis of risk fluctuations be carried out for different enterprises from the same perspective.

We have constructed a risk volatility index for this purpose. The risk volatility index can be traced back to 1993, when the Chicago Board Options Exchange (CBOE) released the world's first volatility index (VIX) as reference benchmark (Whaley (1993), Fleming (1995)), which is used to reflect market sentiment. At present, the European Futures Exchange, Frankfurt Stock Exchange, Tokyo Stock Exchange, etc. have launched the VIX index and issued various financial products with VIX as the core, which has become important for investors to manage risk, asset pricing and trading decisions tool.

The data used in our entire index is derived from stock price data of different companies. The widely observed fact in related research is that the probability distribution of the rate of return will deviate from the random Gaussian distribution, showing the characteristics of spikes and fat tails (Mantegna (1995), Gopikrishnan (1999)). This phenomenon can be observed in financial data at various time scales. The non-Gaussian statistical distribution of price returns makes people believe that price dynamics in financial markets are not a random process, but have their own unique nature.

Therefore, for the Internet industry, especially Internet financial services with greater uncertainty risks, we focus on risk losses under extreme market conditions. To this end, we are inspired by the extreme value theory (Gumbel (2012)) as a means of risk measurement. The extreme value theory mainly includes two types of models: BMM (Block Maxima Method) model and POT (Peaks over Threshold) model.

Among them, the BMM model uses different statistical methods to model and analyze the maximum or minimum values in a series of independent and identically distributed observation data, so it needs to simulate a large amount of data. For Internet finance, the number of listed companies and the time to market are often limited, due to the short development time, and this makes it difficult to obtain a large amount of data.

In contrast, the POT model is more effective than the BMM model. The POT model is less dependent on the amount of data. That is, by selecting the data distribution of all the samples that exceed a certain threshold, the distribution of corporate tail risk can be derived from this distribution.

In the process of POT model, the extreme value method often uses the generalized Pareto distribution. The generalized Pareto distribution method for measuring risk can be traced back to the theory of Pickands, Balkema (1975), and then Larsén (2015) applied the generalized Pareto distribution to the estimation of extreme winds of limited length to investigate the sources of uncertainty in both applications. Dey (2016) apply the generalized Pareto distribution to hurricane damage data to quantify the inferred uncertainty under the extreme regression level of hurricane losses. The research results show that as the time period increases, the uncertainty of model extrapolation increases. The generalized Pareto distribution is applied to the Dow Jones Islamic Market Index, the US S&P 500 Index, and the Asia and Europe S&P Index to compare the risks of traditional stock markets with Islamic stock markets. Studies have shown that for traditional stock markets there may be an upper limit on profits in extreme events (Mwamba (2016)).

These studies not only provide inspiration for our research, but also show that generalized Pareto can be applied to the risk assessment of time series in various fields. Based on this, this paper uses the generalized Pareto distribution model to evaluate the risk of enterprises.

First, according to the classical generalized Pareto model, set the function as

$$F(x;\mu,\sigma,k) = \begin{cases} 1-\left(1-k\frac{x}{\sigma}\right)^{\frac{1}{k}} & k \neq 0 \\ 1-e^{-\frac{x}{\sigma}} & k = 0 \end{cases} \quad (1)$$

where $\sigma$ is scale parameter of distribution, and $k$ is the shape parameter of distribution. $\sigma > 0$ and when $k \leq 0, x \geq 0$; when $k > 0$, $0 < x < \frac{\sigma}{k}$; and when $k = 0$, the distribution is exponential.

Among them, the role of threshold is very important. In the POT model, by setting a threshold in advance, all observed data exceeding this threshold are constituted into a data group, and the data group is taken as the object of modeling and applied to the generalized Pareto distribution to calculate the risk value.

We refer to Roth (2015) to summarize the analysis in two different ways to obtain the threshold. A threshold value is obtained by means of threshold stability graph and hypermean graph based on the observation of the icon. The principle of threshold stability graph is to find a threshold $u_0$. When $u > u_0$, its generalized Pareto distribution function remains unchanged $u = u_0$. The rule of super-mean graph is based on its function $e_n(u) = E[X - U|x > u] = \frac{\sum_{i=1}^{n}(X_i - u)}{n}$, where $i$ is observation of the sample over the threshold, and the graph takes $u$ as the horizontal axis and $e_n(u)$ as the vertical axis to obtain the corresponding function graph. If the function tends to be linear after a certain observation value, it can be determined that the observation value is the required threshold value. The other is to use the goodness of fit test -- KS test, which is a test method to compare the frequency distribution function with the theoretical distribution function or the distribution of two observed values. The test function is defined as: $D_n$=sup$|F_n(x) - G_n(x)|$, and finds the minimum of $a$ that satisfies the condition that $D_n < D(n, a)$, takes $a$ as the quantile of the threshold, takes $n$ as the number of samples, and $a$ as the significance level. Hill proposed a method to calculate the threshold. We take it as Hill estimation method. The independent sample observations of $n$ are arranged in ascending order, which satisfies$x_{(i)} > x_{(i-1)}, i = 2, \dots, n$. According to the estimator: $h_{k,n} = \frac{1}{k}\sum_{i=1}^{k} ln(X_{n-j+1}) - ln(X_{n-k}), k = 1,2 \dots n-1$, $k$ is taken as the horizontal axis and $\frac{1}{h_{k,n}}$ is taken as the vertical axis, then the threshold selects $x_k$ corresponding to the abscissa $k$ of the initial point of the stable region in the graph as the threshold $u$.

There are still a few things to note when choosing confidence levels. Generally speaking, the higher the confidence level, the better the extreme value model can capture the risk characteristics of the distribution. Too high confidence level may lead to too little observed data and increase the variance, but if the selected confidence level is too low, the generalized Pareto approximate estimation parameters are not valid, resulting in the estimator of the deviation. Because of the quantity of data in this paper and experiments, the threshold $u$ is finally selected at the confidence level of 80% in this paper.

First, we use the excess rate of return to process the time series data:

$$AR_{\mathrm{i,t}} = R_{\mathrm{i,t}} - R_t = \frac{a_{\mathrm{i,t}} - a_{\mathrm{i,t-1}} - a_{i,t-1}R_t}{a_{i,t-1}} \tag{2}$$

Where $AR_{\mathrm{i,t}}$ is the excess return rate of the company $i$ at time $t$, $R_{\mathrm{i,t}}$ is the stock return rate of the company $i$ at time $t$; $R_t$ is the risk-free rate of the market at time $t$; $a_{\mathrm{i,t}}$ is the stock closing price of the company $i$ at time $t$ .

Secondly, through the maximum likelihood estimation of the density function of generalized Pareto distribution, the excess return rate can estimate the scale parameters $\sigma$ and shape parameters $k$ of generalized Pareto distribution in different time periods.

Assume that X=$(x_1, x_2 \cdots, x_n)$ is the random variable group of generalized Pareto, where $x_{(n)} = \max_{1 \le i \le n}\{x_i\}$, then its density function is:

$$f(X;\mu,\sigma,k) = \begin{cases} (\frac{1}{\sigma})\left(1-k\frac{X}{\sigma}\right)^{\frac{1}{k}-1} & k \ne 0 \\ 1-e^{-\frac{X}{\sigma}} & k=0 \end{cases} \tag{3}$$

Then the logarithm natural function of the sample is:

$$L(X;\mu,\sigma,k) = ln(\prod_{i=1}^{n} f(X;\mu,\sigma,k)) \begin{cases} -n\,ln\,\sigma + (\frac{1}{k}-1)\sum_{i=1}^{n} ln(1-k\frac{X_i}{\sigma}) & k \ne 0 \\ -n\,ln\,\sigma - \frac{1}{\sigma}\sum_{i=1}^{n} X_i & k=0 \end{cases} \tag{4}$$

Where when $k \le 0$，$\sigma > 0$ and when $k > 0$ 时，$\sigma > kX_{(n)}$. Grimshaw S D (1993) proposes that when $k > 1$，there is no maximum likelihood estimate and when $k$=0 ，it ought to satisfy $\forall 1 \le i \le n$ ，${x_i}^2 = 2x_i$ ，therefore $k$=0 can be rejected. So all we should consider is that when $k \le 0$，$\sigma \ge 0$ and $0 < k \le 1$，$\frac{\sigma}{k} > x_{(n)}$ . So here is the process of the maximum likelihood estimate on the function，

$$\begin{cases} \frac{\partial L\ (X,\mu,\sigma,k)}{\partial \sigma} = -\frac{1}{k^2}\sum_{i=1}^{n} ln(1-k\frac{X_i}{\sigma}) - (\frac{1}{k^2}-\frac{1}{k})\sum_{i=1}^{n}(1-k\frac{X_i}{\sigma})^{-1} + \frac{n}{k}(\frac{1}{k}-1) = 0 \\ \frac{\partial L\ (X,\mu,\sigma,k)}{\partial k} = \frac{1}{\sigma}(\frac{1}{k}-1)\sum_{i=1}^{n}(1-k\frac{X_i}{\sigma}) - \frac{n}{k\sigma} = 0 \end{cases} \tag{5}$$

$$\begin{cases} \frac{1}{n}\sum_{i=1}^{n}(1-\frac{\hat{k}}{\hat{\sigma}}X_i)^{-1} - (1+\frac{1}{n}\sum_{i=1}^{n} ln(1-\hat{k}\frac{X_i}{\hat{\sigma}}))^{-1} = 0 & (6) \\ \hat{k} = -\frac{1}{n}\sum_{i=1}^{n} ln(1-\hat{k}\frac{X_i}{\hat{\sigma}}) & (7) \end{cases}$$

According to formula (7), it can be converted from binary parameter estimation to single element estimation and that is, as long as $\sigma$ is estimated, $k$ can be estimated through formula (7). Because of the formula (6) (7) is based on the $\frac{\hat{k}}{\hat{\sigma}}$ closed form, so that let $\frac{\hat{k}}{\hat{\sigma}}=\hat{b}$, the estimating $k$ and $\sigma$ into $b$ and $k$. Then the above formula (6) and (7) can be translated into:

$$
\begin{cases}
\frac{1}{n}\sum_{i=1}^{n}(1-\hat{b}X_i)^{-1} - (1+\frac{1}{n}\sum_{i=1}^{n} ln(1-\hat{b}X_i))^{-1} = 0 & (8) \\
\hat{k} = -\frac{1}{n}\sum_{i=1}^{n} ln(1-\hat{b}X_i) & (9)
\end{cases}
$$

Where $b < X_{(n)}^{-1}$, in order to ensure the upper bound convergence of the maximum likelihood function, Zhang (2007) proposed an algorithm to obtain the final estimate of $b$ through continuous numerical iteration, and then obtained the estimate of $k$ and $\sigma$.

Theorem $g(b) = \frac{1}{n}\sum_{i=1}^{n}(1-bX_i)^p - (1-r)^{-1} = 0$, where $p = \frac{rn}{\sum_{i=1}^{n} ln(1-bX_i)}$ and $r < 1$ satisfies:

(1) $g(b)$ is a smooth monotone function of $b$, unless $r = 0$ or $X_1 = X_2 = \cdots = X_3$

(2) $when\, r < \frac{1}{2}$, $r \neq 0, n > 2$, $\lim_{b\to-\infty} g(b) < 0$ $and$ $\lim_{b\to (X_{(n)}^{-1})} g(b) > 0$

By iterating through the values, a unique estimate $\hat{b}$ can be found, and the estimate $\sigma$ can be obtained through $\hat{\sigma} = \frac{\hat{k}}{\hat{b}}$.

Since the scale parameters $\sigma$ and shape parameters $k$ in generalized Pareto distribution are determined by the maximum likelihood method, the maximum likelihood estimate $\hat{\sigma}$ and $\hat{k}$ is obtained. According to the previous derivation, setting the time period $r$, the VaR calculation formula is

$$
\mathrm{VaR}_{i,m} = \mu + \hat{b}\left[\frac{n}{N_u}(1-p)^{\frac{1}{n}\sum_{r=m'}^{m''} ln(1-\hat{b}\frac{a_{i,r}-a_{i,r-1}-a_{i,r-1}R_t}{a_{i,r-1}})} -1\right] \quad (10)
$$

where $\mu$ is the threshold, $\hat{b}$ is the estimate, $n$ is the total number of samples, $N_u$ is the number of samples exceeding the threshold $\mu$, $P$ is the confidence level selected, and then $VaR_{i,m}$ is the risk value of the company $i$ in the month $m$. $R_{i,r}$ is the stock return of the company $i$ at time $r$, $R_t$ is the risk-free rate in the market at time $t$, $a_{i,r}$ is the stock closing price of the company $i$ at time $r$. $r$ and $r-1$ time points are included in the month $m$.

$m'$ is the start of the month $m$. $m''$ is the end of the month $m$.

On this basis, the rise and fall of the risk value in a specified period of time can be converted into the way of slope. The positive or negative slope indicates whether the risk increases or decreases in the corresponding time series. That is:

$$RFR = \frac{\hat{b}\frac{n}{N_u}\left[(1-p)^{\frac{1}{n}\sum_{r=(m+1)'}^{(m+1)''} ln(1-\hat{b}\frac{a_{i,r}-a_{i,r-1}-a_{i,r-1}R_t}{a_{i,r-1}})} - (1-p)^{\frac{1}{n}\sum_{r=m'}^{m''} ln(1-\hat{b}\frac{a_{i,r}-a_{i,r-1}-a_{i,r-1}R_t}{a_{i,r-1}})}\right]}{t} \tag{11}$$

# VII. Assessment strategies

In this section, we take four representative Internet technology companies -- Facebook, PayPal, Alibaba and JD -- as examples and list them as a set of comparative data to explain our analysis strategy. It should be noted that all four companies operate or invest in businesses with Internet finance. However, since some of the invested companies are not listed on the stock market, it is difficult to obtain the stock price data, so we choose to use their parent companies, namely the four companies mentioned above, as the preliminary research objects. The main purpose of this section is to explore the effectiveness of the analysis strategy to lay the foundation for the comparative analysis in the following article.

First, we used the original risk model to analyze the risk volume of the four companies. Then, we used the RFR method to process the original data of the four companies. After that, we used the accumulation area chart to represent the risk proportion of each company and the risk change trend of the whole data set. Since the value of the stacking area graph is expressed by relative height, the stack area map will not be covered or hidden with data points of different categories.

The following is our whole comparative analysis process:

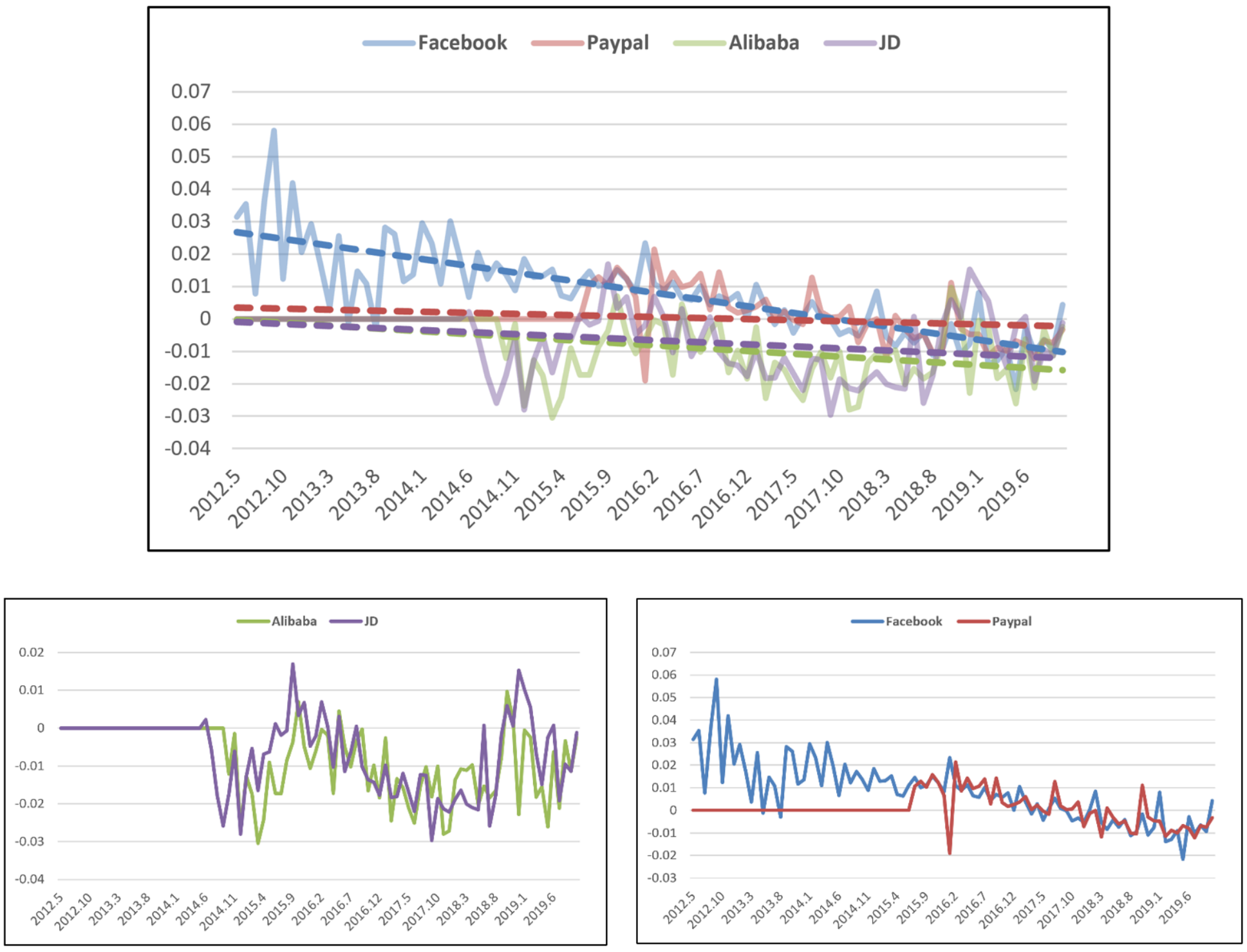

**Figure 2. The risk volume under the original risk model.** The figure 2 shows the risk volume, the bottom left picture shows the risk comparison between Alibaba and JD, and the bottom right picture shows the risk comparison between Facebook and Paypal. The horizontal line in the figure represents the period of time that has not yet been available.

As can be seen from the above image, Facebook was the earliest to go public; the risk was higher in the early stage, and gradually decreased in the later stage. The results show Paypal had the highest risk, followed by Facebook, JD and Alibaba.

It is, therefore, just as we feared; such original risk volume cannot describe the risk contrast well. For example, the trends of Alibaba and JD almost coincide, and so do the trends of Facebook and Paypal.

We believe this result is due to the market in which the main business is located. Alibaba and JD's main businesses are in China, while Facebook and Paypal's main businesses are in the United States. Influenced by the regulatory constraints and market environment in the two countries, the trend is almost similar. Therefore, in the following analysis, we will consider the impact of the overall market environment and macroeconomic development on the risk trend judgment. According to the stock market where the enterprise is listed and the important stock market of the country where its main business is located, we join IXIC, DJIA, S&P; Shanghai Composite Index, Shenzhen Component Index and Hang Seng Index in the later stage, in order to carry out a fuller comparative analysis.

To test our guess at market correlation, we did further analysis. First, regulators' focus is usually long, and the fact that these four companies are by themselves does not suggest that different markets determine risk dynamics. To overcome these limitations, we measured the correlation between the risk volume index and the RFR index for all enterprises.

We use Pearson linear correlation coefficient, which assumes that the covariance between two random variables is:

$$COV(\xi,\eta) = E(\xi - E\xi)(\eta - E\eta)$$

Pearson correlation coefficient is defined as:

$$\rho_{\xi\eta} = \frac{Cov(\xi,\eta)}{\sqrt{D\xi}\sqrt{D\eta}}$$

Where $\rho_{\xi\eta}$ is a dimensionless quantity, and $-1 \le \rho_{\xi\eta} \le 1$. $\rho_{\xi\eta} = 1$ and $\rho_{\xi\eta} = -1$ represent two random variables that are positively and negatively correlated.

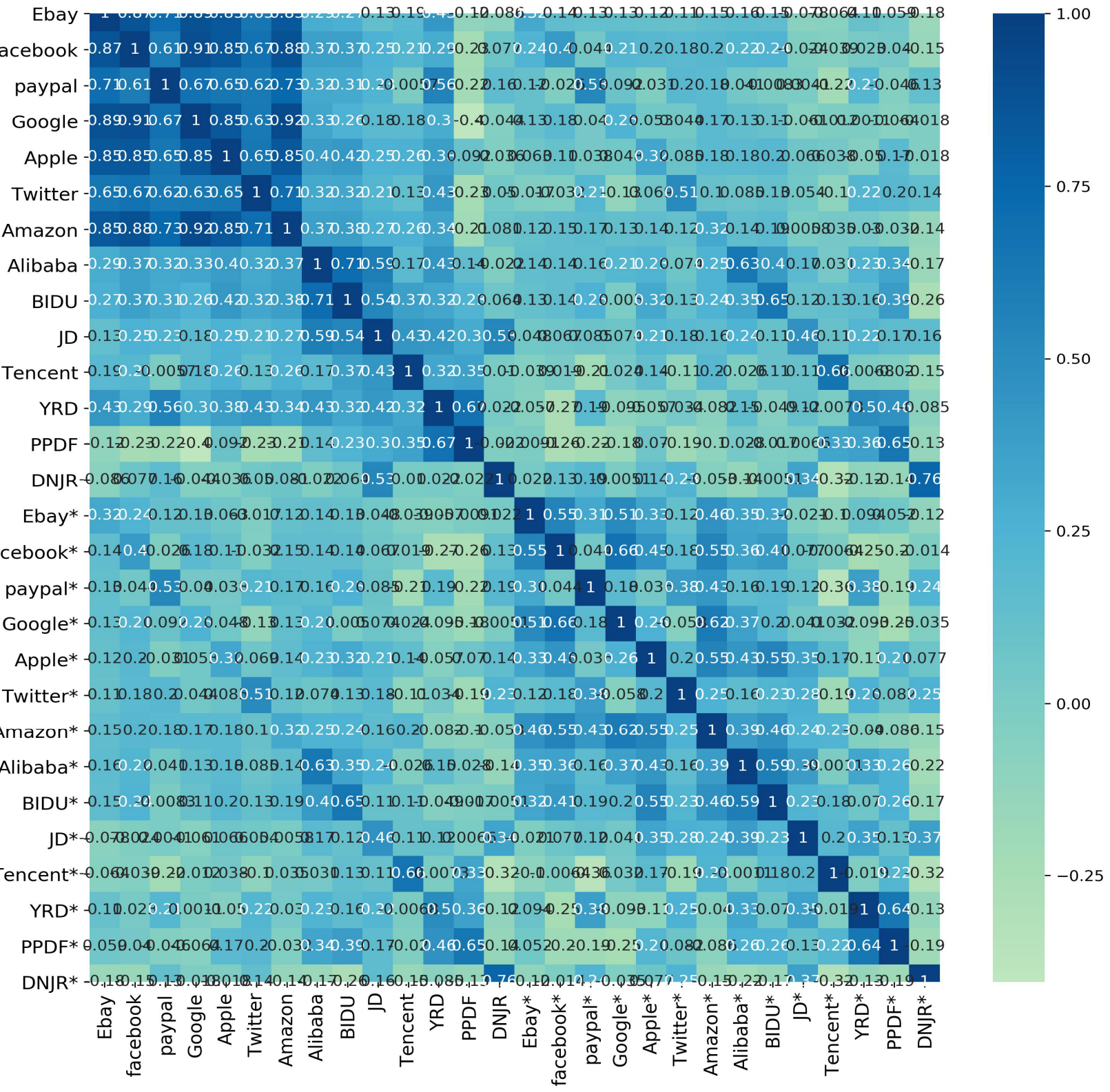

**Figure 3. Correlation between risk mass index and RFR index,** where * denotes the RFR measure value, and no * denotes the risk volume value.

It can be clearly seen from the image that enterprises in the US market have the highest correlation in terms of risk volume. There is almost no such thing in the RFR index. While Pearson linear correlation coefficient is widely used in time series, but it is not a correlation measure robustness, especially under the condition of extreme value will be failure, this is because there is heterogeneity and the stability of data in reality (Zebende (2011)). Therefore, we also need to carry out risk analysis on RFR's local capabilities and conduct more comparative experiments.

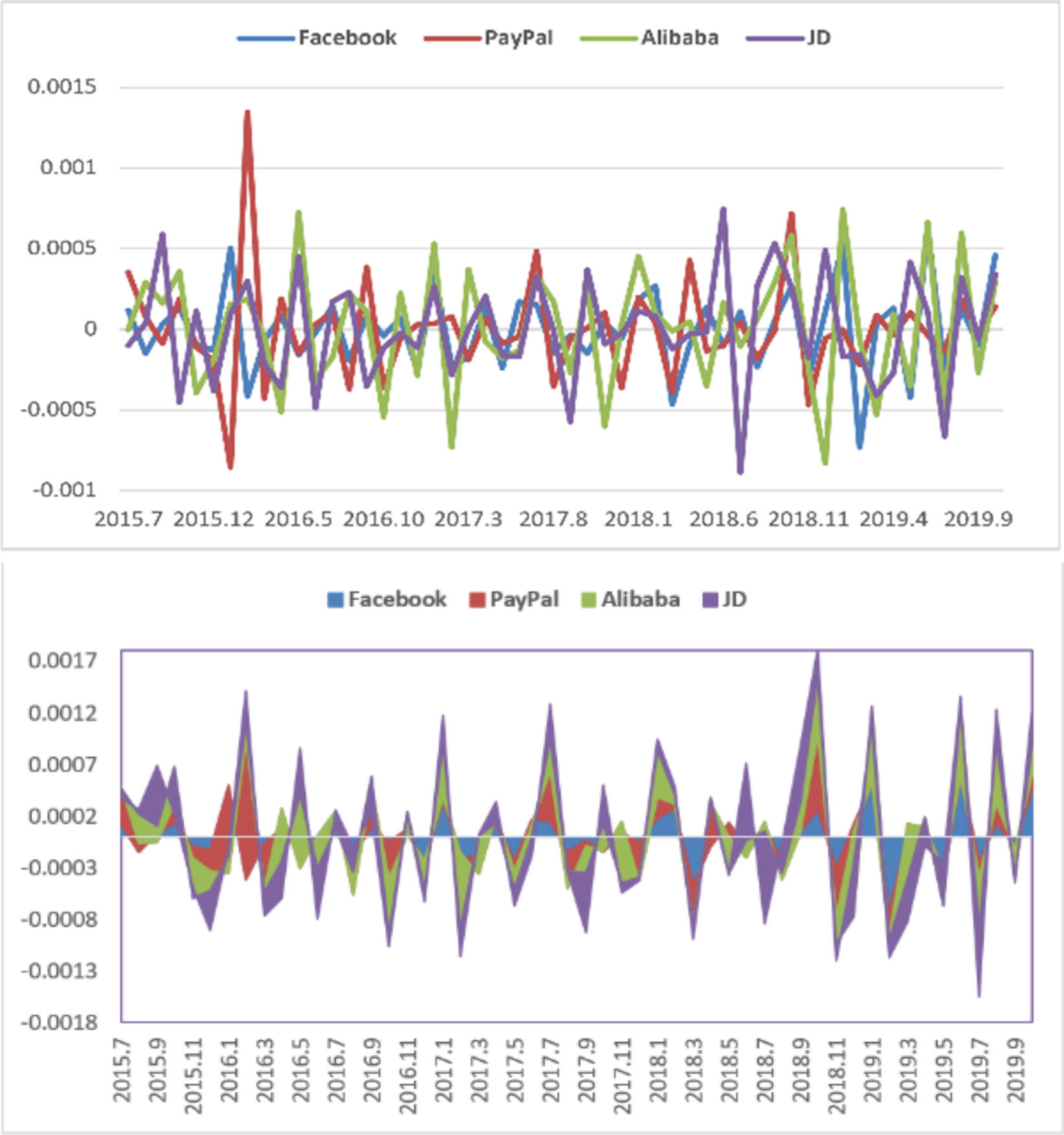

**Figure 4. Comparison of four companies RFR.** The image on the left is the RFR image, and the image on the right is the cascade area diagram of RFR.

A single Internet company can be seen on the left as generating more volatility than others at any given moment. The chart on the right shows the total risk volatility of the four companies. Since the interference of risk volume in the risk analysis process is excluded, we clearly describe the risk size of four different enterprises. We can compare the risk size of four enterprises at different times by the ordinate data of the picture. In addition, the cyclical correlation of the four enterprises also deserves our attention, which will be discussed more fully in the following sections.

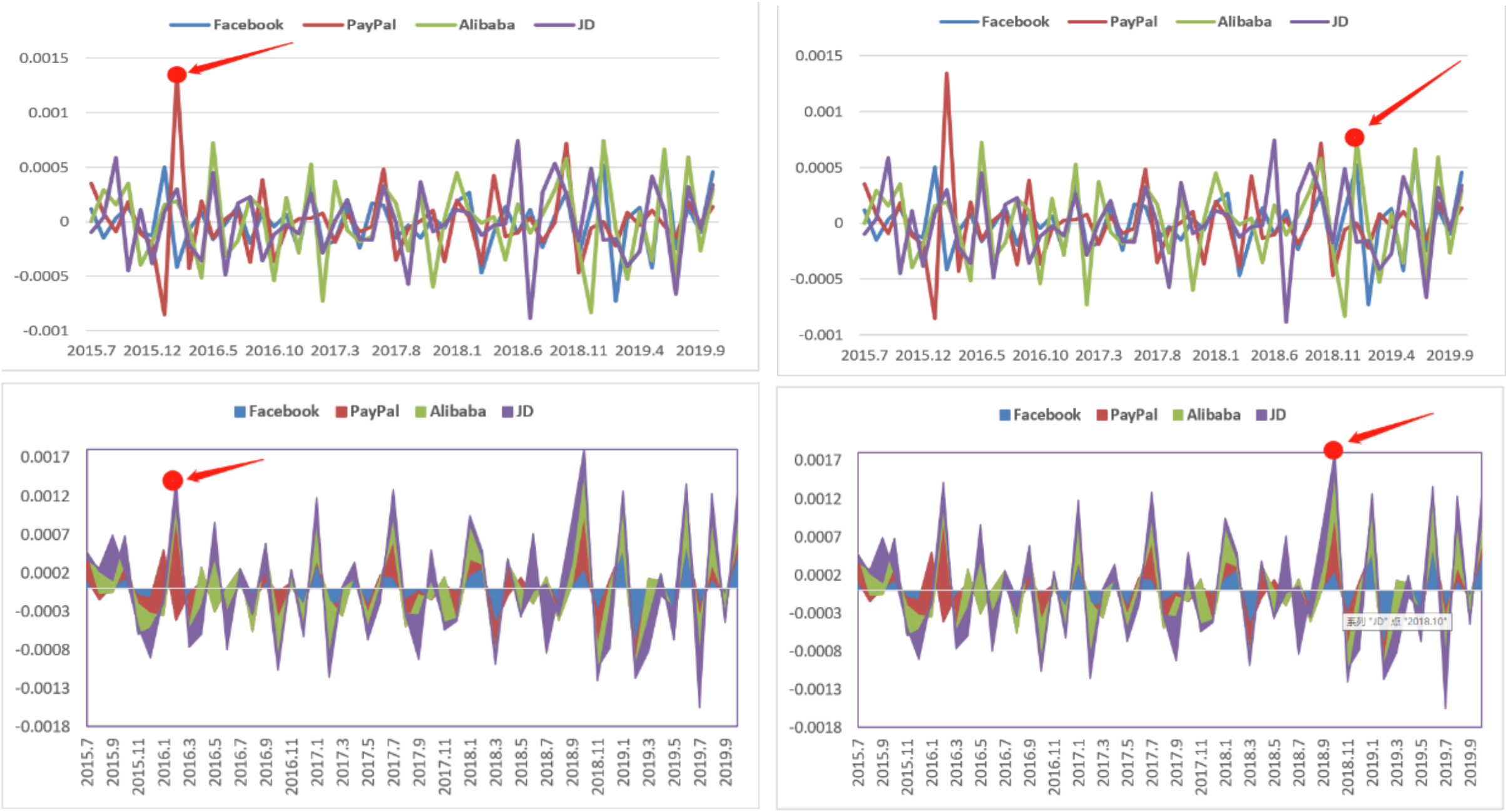

**Figure 5. Comparison of RFR image and area stacking image**

If individual companies don't show very substantial risk growth, every company does show some degree of risk growth. Then the figure on the right will produce a large risk area, indicating that the enterprise risk in the industry is rising together, and the industry as a whole tends to increase risk. After the occurrence of risk in one enterprise, the degree of risk contagion and the success rate of contagion in the industry are reflected in the risk index of other companies. When one company has a risk, it can produce effective risk contagion to other companies. In the specified period of time, the risk of the industry will rise together. If the process of risk contagion is blocked due to good risk control mechanism in the industry, only the company will take risks, and the total risk of the whole industry tends to be stable.

The advantages of this method are that on the one hand, it combines macro- and micro-analysis; on the other hand, it considers the influence of risk in the process of infection. In addition, according to the selection of threshold value, we can use day, week, month and so on as units to measure, and can flexibly transform the required observation index details.

# VIII. Risk Correlation under RFR Approach

With the mutual influence and infiltration of financial and economic activities, as well as the massive transmission and exchange of market information, the interaction, behaviour and mutual correlation among financial markets also show a significantly rising trend. The interactive behaviour of the financial market promotes the optimal allocation of financial and economic resources, but also leads to the frequent outbreak of risk in recent years. This is due to the fact that economic and financial development in all countries of the world are closely connected, whether by each country's financial markets of the global financial system, or a country. Even in the financial system, there are indivisible and complex relationships among many financial individuals, and they eventually constitute the complex financial systems of various sizes.

Internet technology companies are similar in business model, operation model, profit model, user characteristics, etc. Therefore, we suspect that their risk fluctuation model is also correlated to some extent.

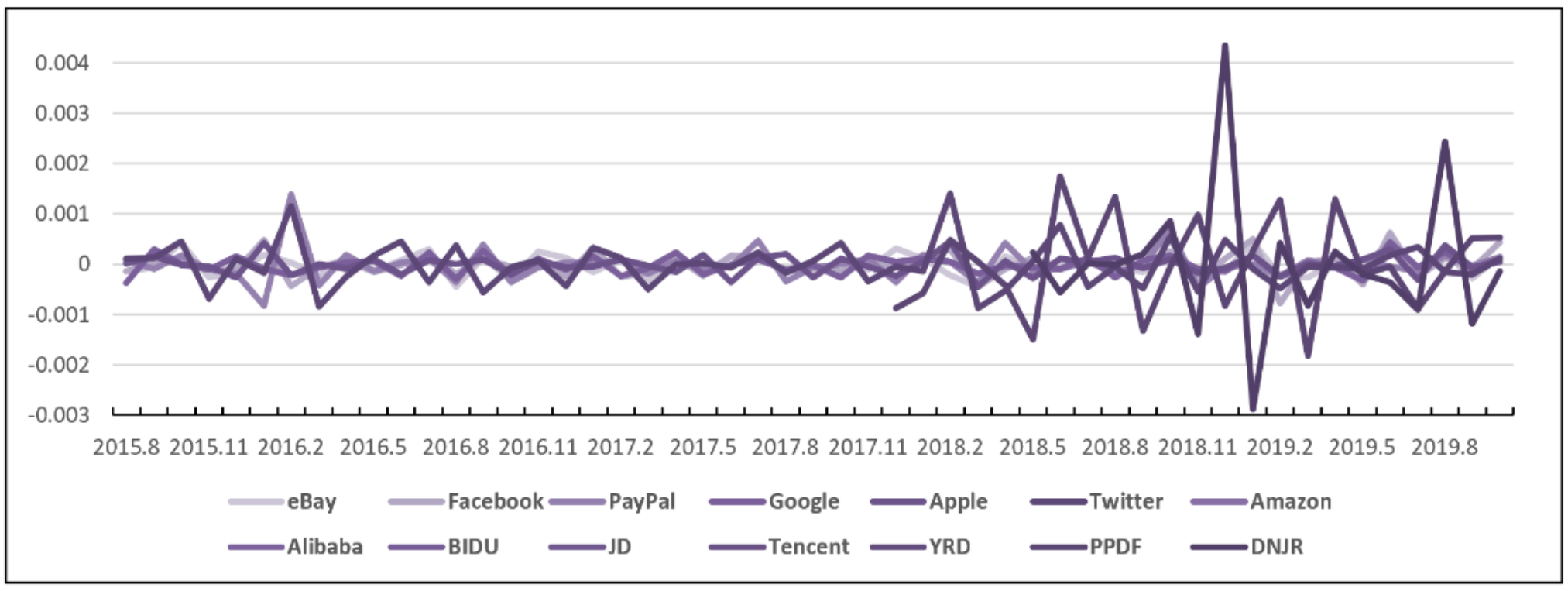

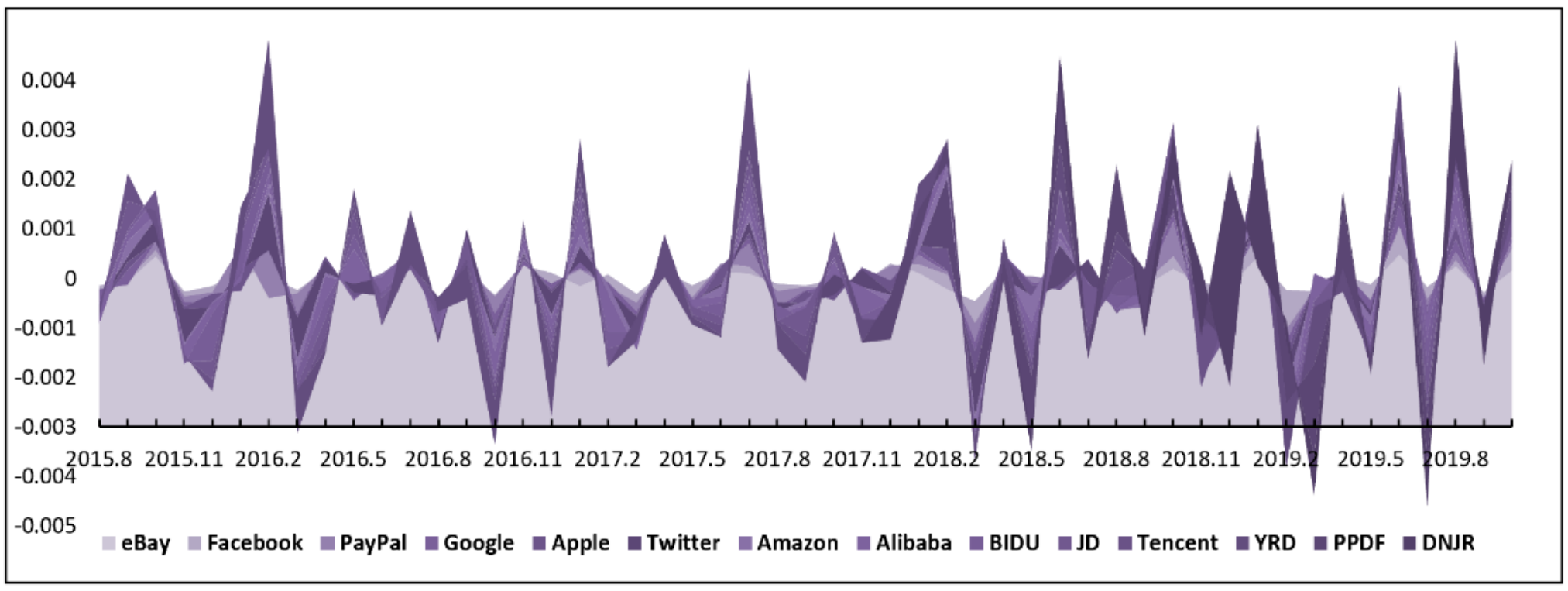

**Figure 6. Risk fluctuation amplitude analysis of 14 companies under the RFR method.**

The figure 6 shows the risk fluctuation amplitude of fourteen companies under the RFR method. After observation, it can be found that the companies that generated huge amplitude after time series 2018.2 are Internet finance companies, namely YRD, PPDF and DNJR.

The following picture considers the risk accumulation area after Internet finance companies. After observation, it can be found that the dark part after time series 2018.2 represents the fact that Internet finance companies deviate from the overall industry trend and generate huge risk fluctuations.

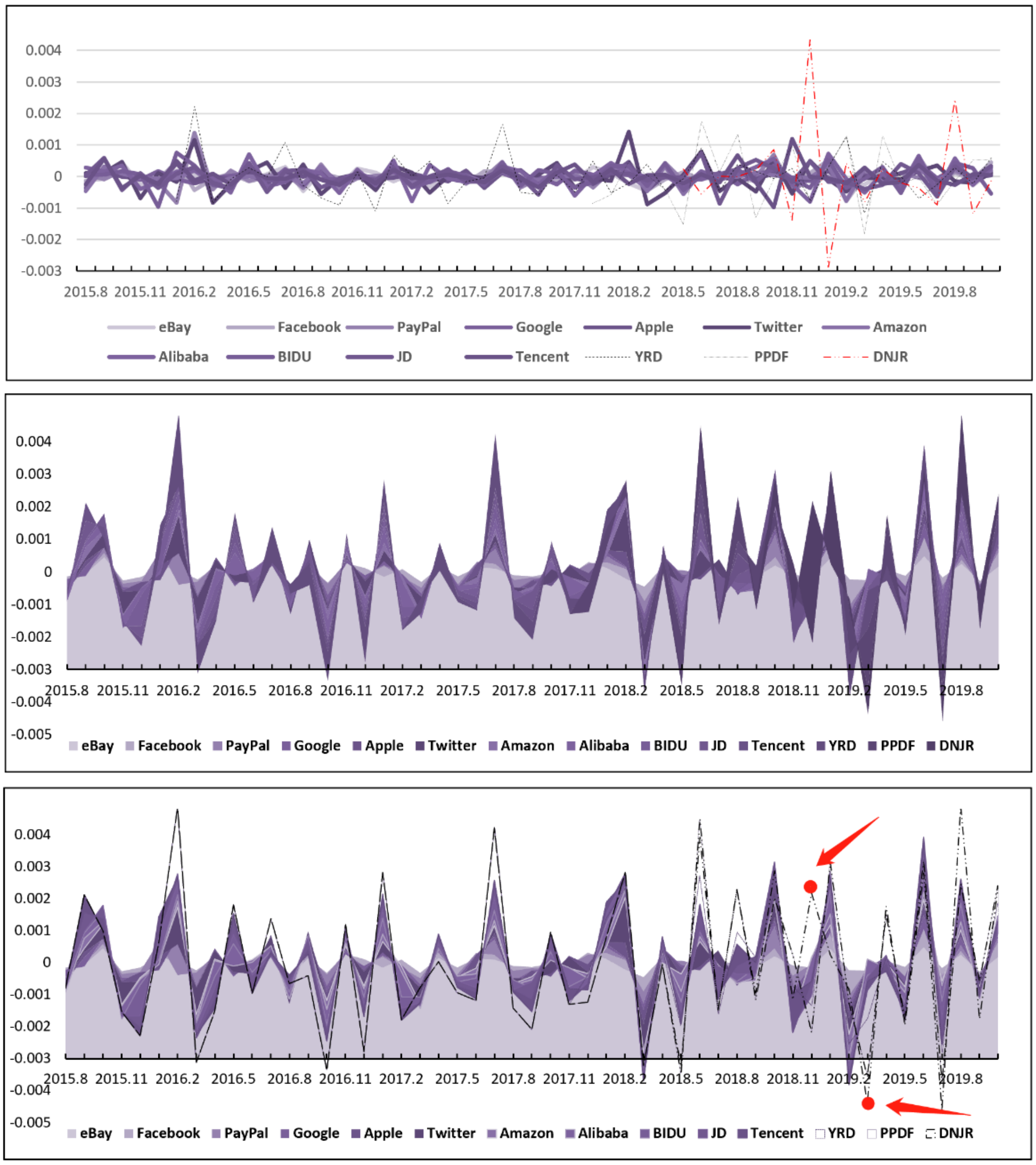

**Figure 7. The influence of three Internet financial enterprises on risk assessment.**

The figure 7 shows the risk volatility of fourteen companies under the RFR method. The red dotted line is the Internet financial enterprise DNJR, which has repeatedly occurred risk events after listing. The other two dotted lines represent Internet finance companies and risk volatility is far greater than other Internet technology companies in the industry. The middle picture shows the risk accumulation area after considering Internet finance companies. The picture below shows the risk accumulation area after the Internet finance company is not considered. The Internet finance companies represented by the dotted line not only deviate from the cycle in risk volatility, but also far higher than other technology companies in risk level.

We explained the reasons for choosing these three Internet financial enterprises before. Because a large number of technology companies invest or directly engage in Internet finance-related businesses, but the number of listed companies is very small, so we can only choose the parent company as the analysis object. YRD, PPDF and DNJR are pure Internet financial enterprises, which are more representative in the analysis.

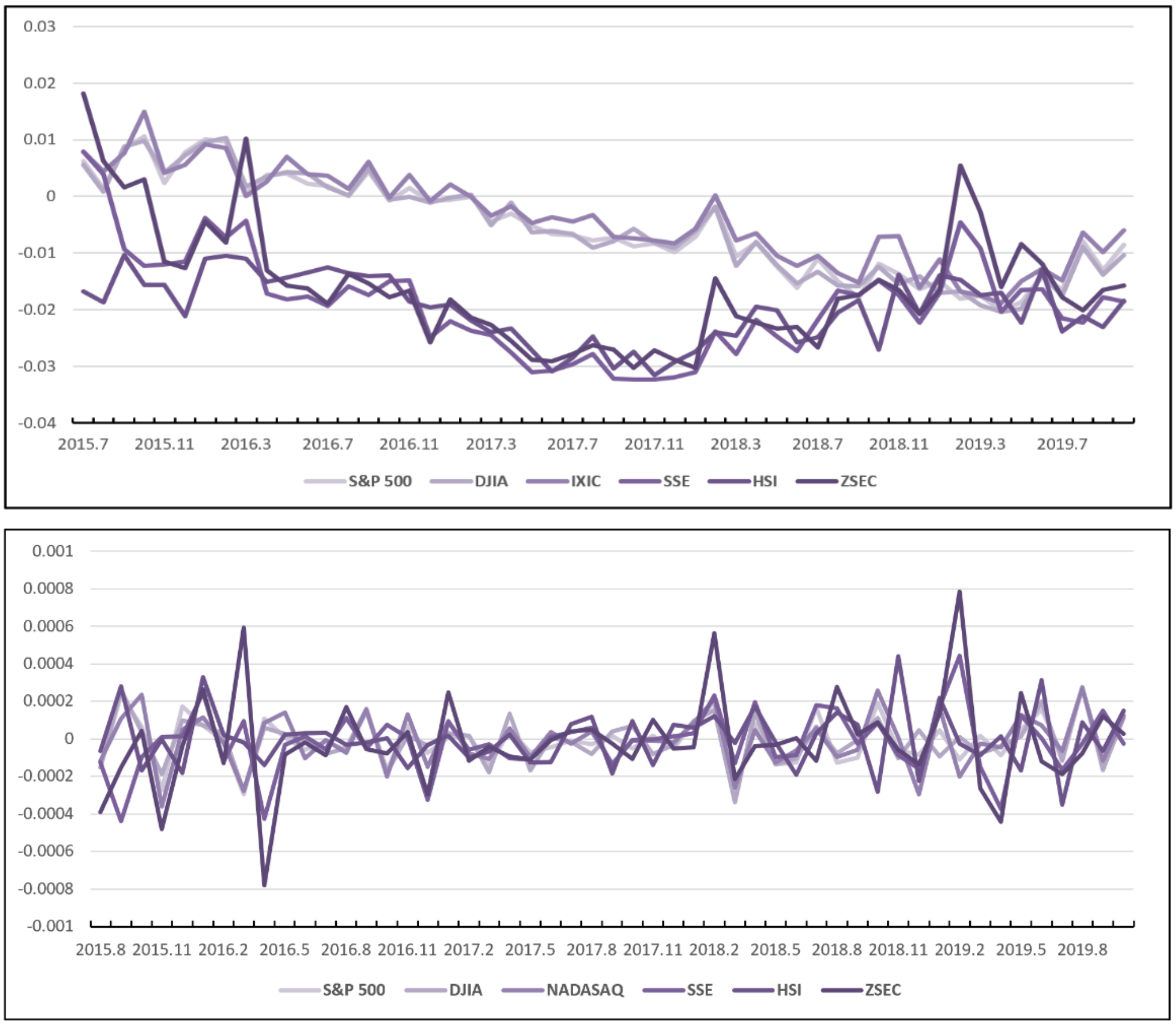

**Figure 8. Risk assessment of composite index.**

The fiugre 8 shows the risk volume of various indexes, and the picture below shows the risk fluctuation range of various indexes under the RFR method.

Since the time series value of RFR is determined by both the previous time series and the fluctuation range, it therefore follows that the RFR method is better for comparative analysis, which can be used to judge the overall trend of a set of data and the differences in details in the trend process. In addition, we can find that the correlation of various indexes on risk periodicity is also different. Measured by the value at risk, the overall trend is broadly similar. However, in the case of risk volatility, there is no strong cyclical correlation between technology companies.

Figure 8 about here

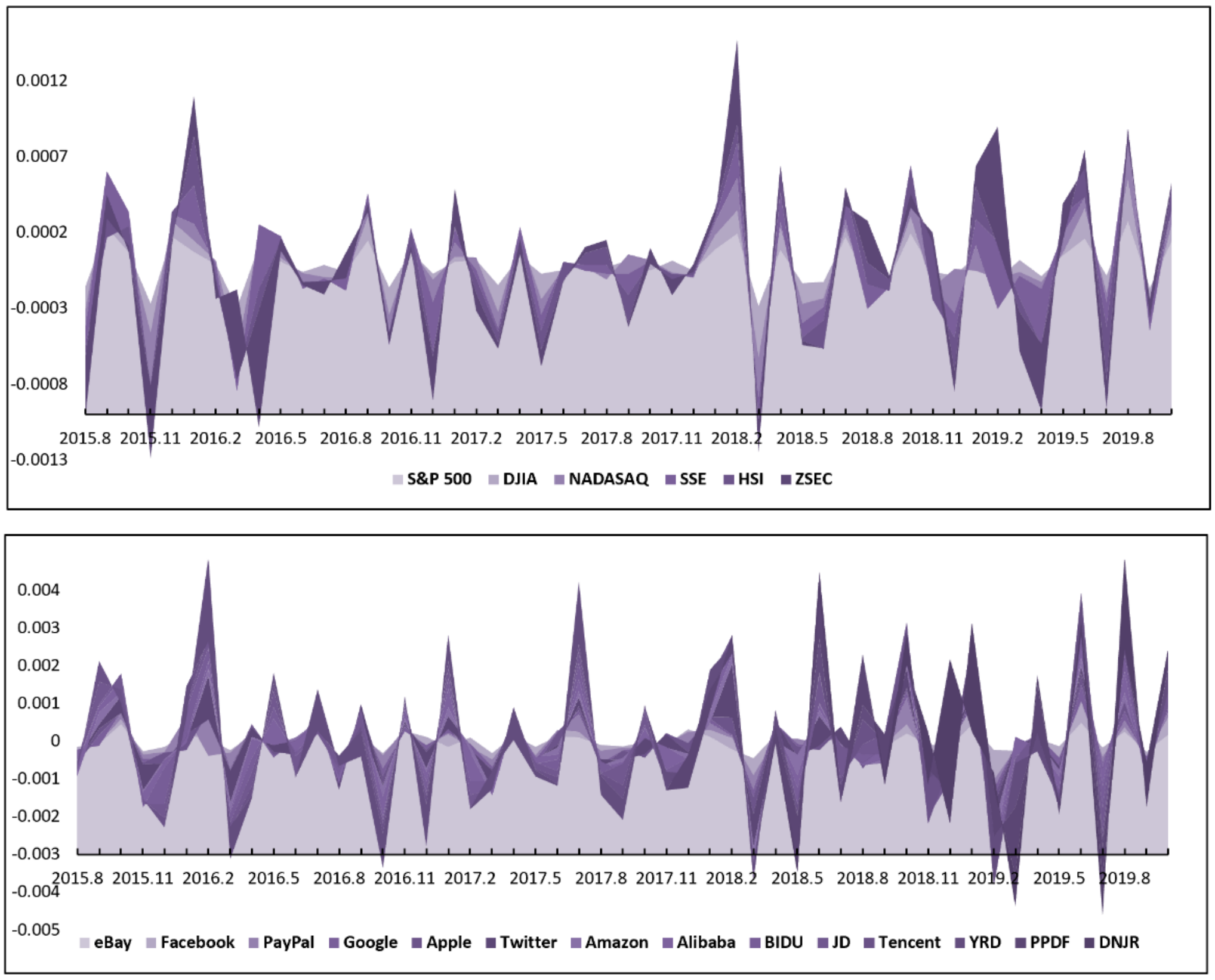

**Figure 9. Comparative analysis of risk accumulation area between composite index and technology company.**

The figure 9 shows the risk accumulation area of the composite index, and the picture below shows the risk accumulation area of fourteen companies. From comparisons, it can be found that the Internet technology industry and the trend of the composite index is significantly different. Thus, it can be proved that the strong correlation of cycles is a conjecture formed due to the special characteristics of the Internet industry. In addition, the risk volatility of the composite index is much smaller than that of the technology industry, as can be seen from the vertical value of the two graphs.

Due to the large number of Internet technology companies listed on Nasdaq, the Nasdaq is an emerging high-tech index that covers companies from telecoms to biotechnology and is the world's pre-eminent index of large capital growth. We have to consider the Nasdaq because of its huge exposure to the technology sector.

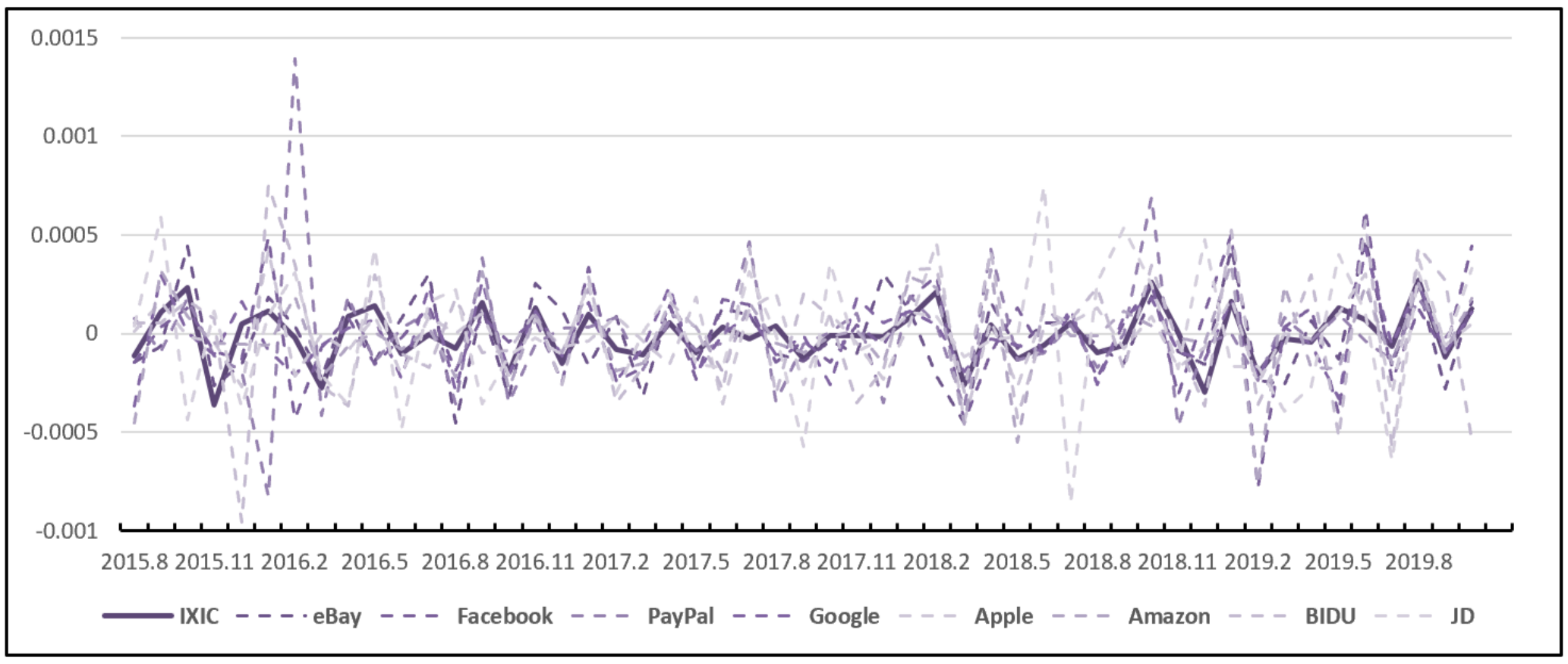

**Figure 10. NASDAQ versus the Risk of Technology Stocks in its Market.**

Obviously, the risk volatility for tech companies is much larger, but the risk trends are almost similar across data values. This proves our conjecture that the securities market where enterprises are listed will lead to the convergence effect of risk changes.

In addition, this also confirms that technology companies do not rule out the possibility of having a higher level of risk control. For example, we can find in the figure that after the risk fluctuates to a very high value, the next time series is a plummeting risk value. There are very few instances of continuous risk escalation.

For this purpose, we counted the number of times when RFR was above and below 0 value, which was used to measure whether the interval of risk trajectory was high risk or low risk in a certain time range.

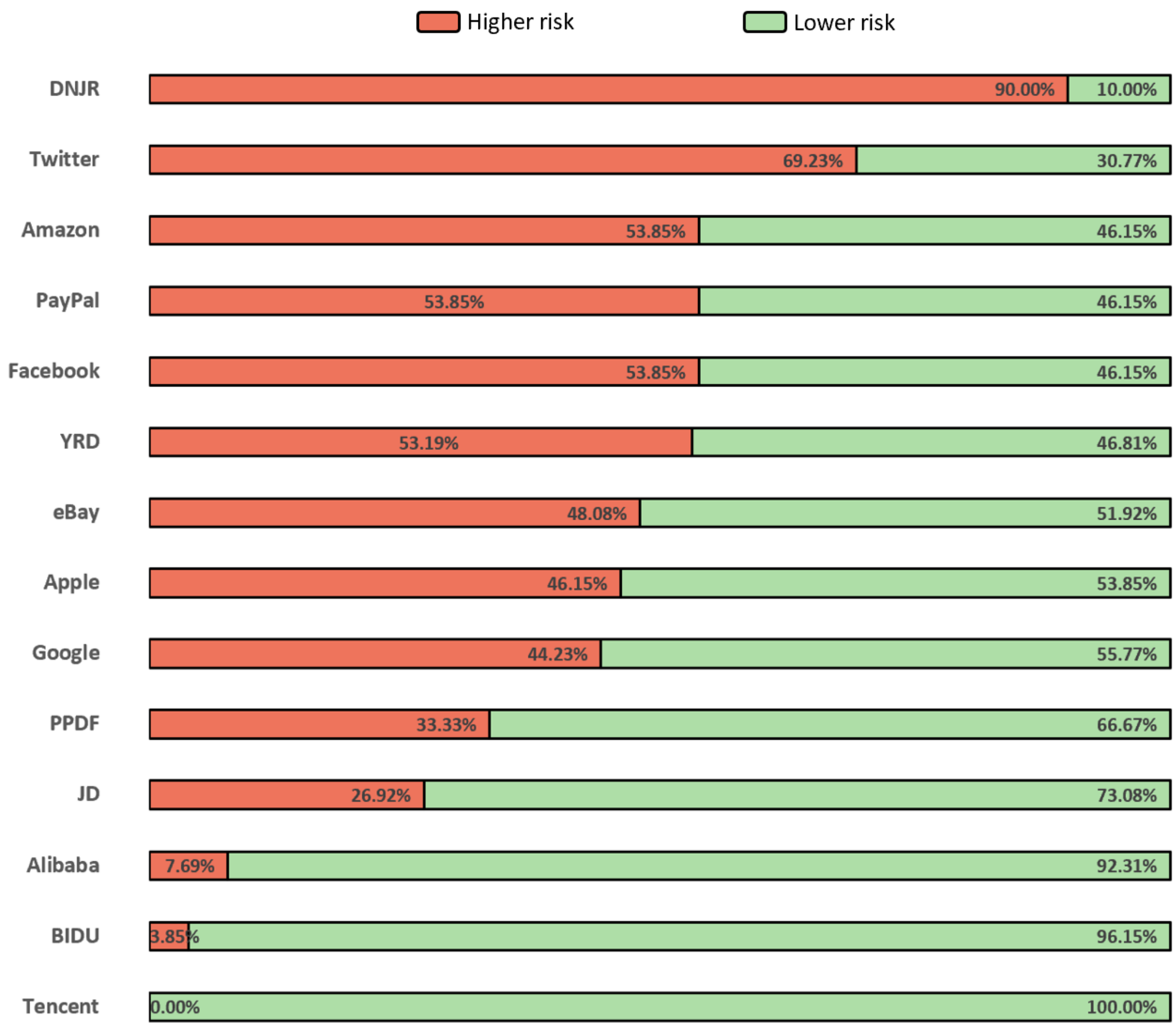

**Figure 11. The percentage of Internet technology companies above and below the RFR value of 0**

The riskiest company in the rankings is DNJR, an Internet finance company that has been delisted because of late payments. DNJR is a Chinese Internet finance company, while all other Chinese tech companies tend to be at the bottom of the list. This has to do with the intensity of regulation in Chinese and American markets. According to table (2), the acquisition behaviours of the 14 companies from 2015 to 2019, it seems that the more acquisitive the company, the better the risk control level remained.

The main reason for the overdue payment of the Internet financial company DNJR is that due to online reports, individual borrowing channels and borrowing customers are fortunate and maliciously overdue, causing the lender's funds to fail to arrive in time. This has been revealed in our previous research (Xu (2020)). The root cause of this kind of risk is that Internet finance companies themselves are pursuing user traffic, which has reduced customer screening conditions. The more serious behaviour is that the Internet financial company dividends user loans and borrowings in the form of subsidies and rewards, which has continuously reduced the quality of the user base.

In this regard, we have a potential concern. For technology industries with frequent acquisitions and investments, the important source of risk is the "external effects" of microeconomic risk-taking activity, which is a single company. The losses imposed on the society by the major risks of the institution are far greater than the losses suffered by investors themselves. In practical terms, powerful companies often pay great attention to their own risk prevention and control, but ordinary platforms that have direct business dealings with these large platforms do not pay enough attention to their own risks. If a powerful company acquires Internet financial companies only to improve its position in a new field or for its own ecological construction, then Internet financial companies are very likely to risk.

The giants in the Internet industry monopolize more user traffic in the market, and will make concessions, subsidies, and increase investment for more traffic. In the field of Internet finance, this behaviour is a risk. From the perspective of the nature of investors in Internet financial products, if most investors participating in Internet financial products become users driven by benefits, it is likely that some users will lack the knowledge of investment and risk.

If Internet financial enterprises ignore the traditional investment behavior of identity authentication for the number of users, or offer an investment interest higher than the industry average industry to result in evil competition,and users who are lack of knowledge of investment and risk cause confusion to platform, it is likely to cause additional exposure to Internet financial enterprises, and the uncertainty of exposure to more easily overlooked by managers.

According to the reasons for the risks of DNJR, we can find that the screening of user quality and user access has become an important part of the Internet finance field.

# IX. Conclusion

This paper uses the stock price data of Internet finance and Internet technology companies whose business includes Internet finance to construct a risk fluctuation model (RFR). This model can flexibly switch the required observation scales, such as day, week, month, etc., according to demand. We show that this new risk volatility model performs well in comparative risk analysis.

The experimental methods we use are the risk mass method and the risk fluctuation method (RFR). Through a large number of comparative experiments, we found that according to the main business countries of different companies, there will be differences in the movement model of risk mass. Countries with the same main business have a high correlation in the fluctuation pattern of risk volume. We also found that under the RFR method, the risk fluctuation amplitude of the entire Internet industry has a cycle-like characteristic, which is affected by the Nasdaq index. The Nasdaq index has an impact on the trend of the Internet industry in terms of risk volume and risk fluctuation range RFR.

We also discuss the business types of Internet finance and study the frequent acquisitions and investment activities in the entire Internet industry. We explain the purpose of this behavior and analyze its impact on the ecological value and risk exposure of the entire Internet industry.

Finally, we raise some concerns about the Internet industry. For the acquisition and investment of the Internet industry, the important source of risk is the "externality" of microeconomic risk-taking activity. If a large Internet company only considers the future ecological value and does not care about the profitability of the acquired enterprise in the process of acquiring the Internet finance enterprise, then the acquired Internet finance enterprise is prone to risk exposure of crisis events and various uncertainties.

Although a large number of research papers have been devoted to exploring the business models and risk relationships of Internet technology companies over the past two decades, there has been relatively little structured research on the subject (especially theoretical construction work and empirical studies beyond single case studies). What's more, much of the research on business models has been done in the context of start-ups, so we know less about the business models of mature companies – for example, how they change over time. Indeed, while the business model approach may shed new light on the fundamental questions of strategic entrepreneurship, much remains to be explored. This approach can also help us expand and even rethink some of the accepted wisdom in core strategy and entrepreneurship.

# APPENDIX

## Appendix A. Internet finance companies in some countries (both listed and unlisted)

| Company | Country | Key Services | Listing Place |
|---|---|---|---|
| Alibaba | China | Mobile payment | NYSE |
| Tencent | China | Mobile payment | HKEX |
| Lakala | China | Mobile payment | SZSE |
| PNR | China | Mobile payment | HKEX |
| LX | China | Debit and Credit | NASDAQ |
| DNJR | China | Debit and Credit | NASDAQ |
| FINV | China | Debit and Credit | NYSE |
| XYF | China | Debit and Credit | NYSE |
| WEI | China | Debit and Credit | NYSE |
| YRD | China | Debit and Credit | NYSE |
| U51 | China | Debit and Credit | HKEX |
| HX | China | Debit and Credit | NASDAQ |
| AIHS | China | Debit and Credit | NASDAQ |
| XRF | China | Debit and Credit | NYSE |
| QFIN | China | Debit and Credit | NASDAQ |
| UNIS | China | Investment | HKEX |
| HongLing | China | Debit and Credit | |
| LU | China | Debit and Credit | |
| Tenpay | China | Mobile payment | |
| UnionPay | China | Mobile payment | |
| JD Finance | China | Mobile payment | |
| Demohour | China | Crowdfunding | |
| Angelcrunch | China | Crowdfunding | |

| Company | Country | Key Services | Listing Place |
|---|---|---|---|
| Square | America | Mobile payment | NYSE |
| Paypal | America | Mobile payment | NASDAQ |
| LendingClub | America | Debit and Credit | NYSE |
| OnDeck | America | Debit and Credit | NASDAQ |
| LendingFree | America | Debit and Credit | NASDAQ |
| Affirm | America | Debit and Credit | |
| Avant | America | Debit and Credit | |
| Fundera | America | Debit and Credit | |
| Prosper | America | Debit and Credit | |
| Sofi | America | Debit and Credit | |
| Kabbage | America | Debit and Credit | |
| TRE | America | Debit and Credit | |
| CircleUp | America | Crowdfunding | |
| Fundrise | America | Crowdfunding | |
| RealtyMogul | America | Crowdfunding | |
| Ripple | America | Mobile payment | |
| MotifInvestment | America | Investment | |

| **Company** | **Country** | **Key Services** | **Listing Place** |
|---|---|---|---|
| Funding Circle | Britain | Debit and Credit | LSE |
| Transferwise | Britain | Debit and Credit | |
| Neyber | Britain | Debit and Credit | |
| BGL Group | Britain | Debit and Credit | |
| OakNorth | Britain | Debit and Credit | |
| MarketInvoice | Britain | Debit and Credit | |
| iwoca | Britain | Debit and Credit | |
| Zopa | Britain | Debit and Credit | |
| RateSetter | Britain | Debit and Credit | |
| Ratesetter Circle | Britain | Debit and Credit | |
| Azimo | Britain | Mobile payment | |
| RateSetter | Britain | Debit and Credit | |
| Currency Cloud | Britain | Mobile payment | |
| Monzo | Britain | Mobile payment | |
| SumUp | Britain | Mobile payment | |
| WorldRemit | Britain | Mobile payment | |
| Seedrs | Britain | Crowdfunding | |

| Company | Country | Key Services | Listing Place |
|---|---|---|---|
| Credit Saison | Japan | P2P | TSE |
| Money Forward | Japan | Investment | TSE |
| | | | |
| Folio | Japan | Debit and Credit | |
| WealthNavi | Japan | Debit and Credit | |
| Origami | Japan | Mobile payment | |
| LinePay | Japan | Mobile payment | |
| Paidy | Japan | Mobile payment | |
| Money Design | Japan | Investment | |
| One Tap BUY | Japan | Investment | |
| Zaim | Japan | Investment | |

| Company | Country | Key Services | Listing Place |
|---|---|---|---|
| Wirecard AG | Germany | Mobile payment | FWB |
| Money Forward | Japan | Investment | TSE |
| | | | |
| N26 | Germany | Investment | |
| Smava | Germany | Debit and Credit | |
| Auxmoney | Germany | Debit and Credit | |
| Bitbond | Germany | Debit and Credit | |
| liberis | Germany | Debit and Credit | |
| Kreditch | Germany | Debit and Credit | |
| Zencap | Germany | Debit and Credit | |
| Payever | Germany | Mobile payment | |
| gastrofix | Germany | Mobile payment | |
| Optile | Germany | Mobile payment | |
| Payworks | Germany | Mobile payment | |
| Kapilendo | Germany | Crowdfunding | |

| Company | Country | Key Services | Listing Place |
|---|---|---|---|
| PagSeguro Digital | Brazil | Mobile payment | NYSE |
| StoneCo | Brazil | Mobile payment | NASDAQ |
| XP | Brazil | Investment | NASDAQ |
| | | | |
| MercadoPago | Brazil | Mobile payment | |
| PAGGI | Brazil | Mobile payment | |
| VINDI | Brazil | Mobile payment | |
| Creditas | Brazil | Debit and Credit | |
| BizCapital | Brazil | Debit and Credit | |
| Nexoos | Brazil | Debit and Credit | |
| EASYCRÉDITO | Brazil | Debit and Credit | |
| Partyou | Brazil | Crowdfunding | |
| Benfeitoria | Brazil | Crowdfunding | |
| Vakinha | Brazil | Crowdfunding | |
| Doare | Brazil | Crowdfunding | |
| URBE.ME | Brazil | Crowdfunding | |
| Organizze | Brazil | Investment | |
| Bling | Brazil | Investment | |

| Company | Country | Key Services | Listing Place |
|---|---|---|---|
| Paytm | India | Mobile payment | |
| Freecharge | India | Mobile payment | |
| MobiKwik | India | Mobile payment | |
| Faircent | India | Debit and Credit | |
| Lendbox | India | Debit and Credit | |
| i2ifunding | India | Debit and Credit | |
| LenDenClub | India | Debit and Credit | |
| IndiaLends | India | Debit and Credit | |
| Lendingkart | India | Debit and Credit | |
| Capital Float | India | Debit and Credit | |
| Vistaar Finance | India | Debit and Credit | |
| Letsventure | India | Crowdfunding | |
| Catapooolt | India | Crowdfunding | |

| Company | Country | Key Services | Listing Place |
|---|---|---|---|
| Paynow | Singapore | Mobile payment | |
| NETS | Singapore | Mobile payment | |
| InstaReM | Singapore | Mobile payment | |
| FlexM | Singapore | Mobile payment | |
| MatchMove Pay | Singapore | Mobile payment | |
| Funding Societies | Singapore | Debit and Credit | |
| Finaxar | Singapore | Debit and Credit | |
| Validus Capital | Singapore | Debit and Credit | |
| Fastcash | Singapore | Debit and Credit | |
| Smartkarma | Singapore | Investment | |
| **WeInvest** | Singapore | Investment | |
| Mesitis | Singapore | Investment | |
| Vistaar Finance | India | P2P | |
| Letsventure | India | Crowdfunding | |
| Catapooolt | India | Crowdfunding | |

| Company | Country | Key Services | Listing Place |
|---|---|---|---|
| Ayopop | Indonesia | Mobile payment | |
| Brankas | Indonesia | Mobile payment | |
| Cashlez | Indonesia | Mobile payment | |
| CelenganID | Indonesia | Mobile payment | |
| DavestPay | Indonesia | Mobile payment | |
| Bareksa | Indonesia | Investment | |
| Fundnel | Indonesia | Investment | |
| Stockbit | Indonesia | Investment | |
| Xdana | Indonesia | Investment | |
| Kitabisa | Indonesia | Crowdfunding | |
| Aksi Bersama | Indonesia | Crowdfunding | |
| Xedeka | Indonesia | Crowdfunding | |

| | | |
|---|---|---|
| Akseleran | Indonesia | Debit and Credit |
| Amartha | Indonesia | Debit and Credit |
| Ammana | Indonesia | Debit and Credit |
| Cashwagon | Indonesia | Debit and Credit |

## Appendix B. Acquisitions by fourteen companies between 2015 and 2019

| Alibaba (2015-2019) | | |
|---|---|---|
| **Key Businesses** | **Time** | **Key Services** |
| IdsManager | October, 2019 | Provide Cloud Service, Unified Authentication and Access Control for Mobile Applications. |
| Yiupin Gougou | September, 2019 | Provide One-stop Shop to Sell Goods Services |
| Koala | September, 2019 | Provide Overseas Logistics, Cross-border E-commerce, Cross-border Payment, Brand Promotion |
| Teambition | March, 2019 | Provide Cross Team Project Cooperation Services |
| InfinityAR | March, 2019 | Provide Customized AR Equipment and Software Services |
| DataArtisans | January, 2019 | Provide Distributed Systems, Large-scale Data Streaming Services for Enterprises |
| Trendyol Group | August, 2018 | Provide Online Fashion Retail Services |
| Daraz | May, 2018 | Provide Logistics Service, Full Categories of E-commerce Services |
| Sound Connected | May, 2018 | Provide Voice Enhancement, Far-distance Voice Interaction Interface |
| C-SKY Microsystems | April, 2018 | Provide Integrated Circuit Design Dedicated to 32-bit High Performance Low-power Embedded CPU and Chip Architecture License |
| ELEME | April, 2018 | Provide Online Food Delivery Services |
| ZTESoft | February, 2018 | Provide End-to-end BsS/OSS Solutions and Services to Global Telecom Operators, Smart City and Vertical Industry Solutions to Enterprises and Governments. |
| EJOY | September, 2017 | Provide Independent or Cooperative Game Development, Agency Operations Services |
| Damai | March, 2017 | Provide Live Entertainment Ticket Marketing Services |
| Glamours Sales | January, 2017 | Provide Membership-based Online Luxury Retail Services |
| Wandoujia | July, 2016 | Provide Mobile App Store |
| AGTech | March, 2016 | Provide Tutoring Service, Lottery, Sports, Health Club, Movie Show, Ticketing Agent |
| SCMP | March, 2016 | Provide Printing and Publishing Services |
| YouKu | November, 2015 | Provide User Video Sharing Services |
| 365 | August, 2015 | Provide Translation and Interpretation Services for Companies at Home and Abroad |
| Vulnhunt | June, 2015 | Provide Hardware and Software Safety Test Service, APT Defense Technology |
| AdChina | January, 2015 | Provide Online Advertising Technology Development, Transaction Services |

**Amazon (2015-2019)**

| Key Businesses | Time | Key Services |
|---|---|---|
| INLT | September, 2019 | Develop International Transportation Cost Management and Customs Clearance Software |
| IGDB | September, 2019 | Provide API subscription Services |
| E8 Storage | July, 2019 | Construct Storage Hardware |
| Bebo | June, 2019 | Organize E-sports Competitions |
| Sizmek Ad Server and Sizmek DCO | May, 2019 | Provide Online Consumer Statistics, Personalized Advertising Services |
| Canvas Technology | April, 2019 | Provide Intelligent Warehouse Management Services |
| Eero | February, 2019 | Develop Wireless Mesh Router System |
| TSO Logic | January, 2019 | Provide Analytical Solution to Help Customers Optimize and Save Time |
| CloudEndure | January, 2019 | Provide Data Recovery, Continuous Backup, Real-time Migration Development Services |
| Tapzo | August, 2018 | Provide Aggregation of Mobile Applications Services |
| PillPack | June, 2018 | Develop Online Pharmacies to Simplify the Process of Taking Medicine |
| Ring | February, 2018 | Develop and Sell Smart-camera-equipped Doorbells |
| Sqrrl | January, 2018 | Track Cyber Security Threats |
| Blink Home | December, 2017 | Identify and Locate of Cyber Security Threat |
| Dispatch | November, 2017 | Provide Local Delivery Powered by Autonomous Vehicles |
| Goo Technologies | November, 2017 | Construct Online 3D Environments |
| Body Labs | October, 2017 | Develop 3D Body Scanning Technology |
| Wing.ae | September, 2017 | Provide Same-day and Next-day Delivery Services |
| GameSparks | July, 2017 | Construct and Manage Functions in the Game |
| Graphiq | July, 2017 | Collect Details about Products, Places and People |
| Souq.com | July, 2017 | Provide Online Shopping Services |
| Whole Foods Market | June, 2017 | Sell Online High-end Fresh Food |
| Do.com | March, 2017 | Provide Conference Products |
| Thinkbox Software | March, 2017 | Provide Video and Other Media Industries with Design and Content Creation Solutions |
| Harvest.al | January, 2017 | Monitor flagged businesses for data breaches |
| Partpic | November, 2016 | Provide Video Search Engine Services |
| Westland | October, 2016 | Publish and Print |
| Curse | August, 2016 | Construct Game Database and Community |
| Cloud9 IDE | July, 2016 | Develop Integrated Environment that Provides Collaborative Programming for Web and Mobile Developers |
| Emvantage Payments | February, 2016 | Provide Mobile Payment Services |
| NICE | February, 2016 | Optimize and Centralize High Performance Computing and Visualization Workloads |
| COLIS Privé | January, 2016 | Express and Delivery |
| Orbeus | December, 2015 | Develop Image Recognition Based on Neural Network |
| Biba Systems | September, 2015 | Develop and OperateVideo Messaging Applications |
| Safaba Translation Systems | September, 2015 | Provide Automated Text Translation Services |
| Elemental Technologies | September, 2015 | Convert Video Data Format |
| AppThwack | July, 2015 | Design, Develop, and Deploy an Automation Framework Used for Testing all of Intel's Wireless Products. |
| ClusterK | April, 2015 | Offer Software that Enables High Availability in the AWS Spot Market. |
| Shoefitr | April, 2015 | Provide 3D Services to Help Consumers Buy More Suitable Shoes |
| 2lemetry | March, 2015 | Track and Manage IP Accounts and Connect Devices to Enterprise |
| | | Systems |
| Annapuma Labs | January, 2015 | Develop Server Networking Chips |

| Apple (2015-2019) | | |
|---|---|---|
| **Key Businesses** | **Time** | **Key Services** |
| Intel's Smartphone Modem Business | July, 2019 | Manufacture Modems for Smartphones and Non-Smartphones |
| Drive.ai | June, 2019 | Construct Self-driving Cars |
| Stamplay | March, 2019 | Develop Cloud and API Based Back-end Development |
| Platoon | December, 2018 | Release Artists' Works, Customize Marketing Strategies |
| Dialog Semiconductor | October, 2018 | Research and Develop Power Management Chip |
| Shazam | September, 2018 | Provide Mobile Music Recognition Services |
| Akonia Holographics | August, 2018 | Design and Manufacture Augmented Reality Glasses |
| Texture | March, 2018 | Provide E-magazine Subscription Services |
| Tueo Health | December, 2018 | Develop Apps that Monitor Asthma Symptoms while Children Sleep |
| Buddybuild | January, 2018 | Provide Platform for Continuous Integration, Continuous Deployment, and User Feedback |
| Silicon Valley Data Science | January, 2018 | Provide Data Analysis for Large Companies |
| Silk Labs | 2018 | Provide Lightweight AI Technology Support for Consumer Hardware |
| Laserlike | 2018 | Provide Quality Information on Topic across the Web |
| Spektral | December 2017 | Separate People and Objects from the Original Background and |
| | | Overlay New Background |
| Pop Up Archive | December, 2017 | Providee Voice Search Services |
| InVisage Technologies | November, 2017 | Develop Smaller Imaging Techniques that Produce Better Quality Images in a Variety of Non-optimal Lighting Conditions |
| PowerbyProxi | October, 2017 | Develop Modular Wireless Charging and Data Transfer System for High Power Applications |
| Init.ai | October, 2017 | Design Intelligent Assistants for Customer Service Representatives |
| Regaind | September, 2017 | Analyze and Classify Massive Image Data Stream and Assessed Value |
| Vrvana | September, 2017 | Research and Develop AR Helmet |
| SensoMotoric Instruments | June, 2017 | Develop Eye-tracking Hardware and Software Products |
| Lattice Data | May, 2017 | Transform Unstructured "Dark Data" into Structured Data and Analyz by Traditional Methods |
| Beddit | May, 2017 | Develop Sleep Monitoring and Management Equipment |
| Workflow | March, 2017 | Develop an Application that Connects Multiple Apps or Functions of Apps to Form Instructions to Complete Tasks Automatically |
| Indoor.io | December, 2016 | Provide Indoor Mapping Services |
| Tuplejump | September, 2016 | Provide Big Data Analysis and Data Simplification Services |
| Gliimpse | August, 2016 | Build a Secure Online Platform for Users to Manage and Share Private Health Data in the Cloud. |
| Turi | August, 2016 | Develop Recommendation Engines, Sentiment Aalysis, Fraud Detection |
| LegbaCore | February, 2016 | Provide Information Security Consulting Services |
| Flyby Media | January, 2016 | Research and Develo AR Technology |
| LearnSprout | January, 2016 | Track Students' Academic Performances |
| Emotient | January, 2016 | Use Facial Expression Analysis to Determine Person's Emotions |
| Faceshift | November, 2015 | Develop Facial Recognition, Real-time Motion Capture Technology |
| Perceptio | September, 2015 | Help Companies Apply AI to Mobile Phones |
| VocalIQ | September, 2015 | Improve Experience of Speech Recognition |
| Mapsense | September, 2015 | Build tools for Analyzing and Visualizing Location Data |
| Metaio | May, 2015 | Provides Augmented Reality and Computer Vision Solutions |
| Coherent Navigation | April, 2015 | Develop High Integrity Global Positioning System Technology |
| LinX | April, 2015 | Develop and Sell Miniaturized Cameras for Tablets and Smartphones |
| FoundationDB | March, 2015 | Privide Software that Writes 54 Billion Times an Hour |
| Semetric | January, 2015 | Provides Data Collection and Analysis Capabilities to Music Websites |
| Camel Audio | January, 2015 | Develop Synthesizer Plug-in, Effect Plug-in |

| **Facebook (2015-2019)** | | |
|---|---|---|
| **Key Businesses** | **Time** | **Key Services** |
| PlayGiga | December, 2019 | Provide VR and Cloud Gaming Services |
| Beat Games | November, 2019 | Develop VR Games |
| Packagd | September, 2019 | Prvide Live Shopping Services |
| CTRL-labs | September, 2019 | Sell Wristband that Receives Brain Signals |
| Servicefriend | September, 2019 | Build Robots to Help Customer Service Teams |
| GrokStyle | February, 2019 | Provide Visual Search for Product Pictures |
| Chainspace | February, 2019 | Use Blockchain Technology to Facilitate Payments and other Services |
| Dreambit | November, 2018 | Provides Personalized Image Search Engine |
| Vidpresso | August, 2018 | Provides Interactive Elements Tools for TV Broadcasters and Online Video Creators |
| Redkix | July, 2018 | Launch Office Software for Communication and Work Coordination |
| Bloomsbury AI | July, 2018 | Use Natural Language Technology to Allow Machines to Answer Questions about the Information |
| Confirm | January, 2018 | Prvide API to Identify Authenticity of Identification |
| Tbh | October, 2017 | Provide Teenagers with Anonymous Votes |
| Fayteq AG | August, 2017 | Use Software Plug-in for Video Editing |
| Ozlo | July, 2017 | Develop Virtual AI Assistant for Android and iOS |
| Zurich Eye | November, 2016 | Develop Robots that Can Navigate Independently |
| Faciometrics | November, 2016 | Provide Face Image Data Analysis Service |
| CrowdTangle | November, 2016 | Track How Content is Spread on Social Networks for Publishers |
| Infiniled | October, 2016 | Develop LED Light Source Modules |
| Nascent Objects | September, 2016 | Provide Modular Consumer Electronics Platforms that Prototype Products Using Small Circuit Boards, 3D Printing, and Modular Designs |
| Two Big Ears | May, 2016 | Create Spatial Sound Service for Virtual Reality |
| MSQRD (Masquerade) | March, 2016 | Provide Dynamic Video Filter Application |
| Pebbles | July, 2015 | Provide Gesture Recognition Control Technology Services |
| Endaga | October, 2015 | Create Independent Cellular Networks for Rural Communities. |
| Surreal Vision | May, 2015 | Provide Real-time 3D Scene Reproduction Services |
| TheFind, Inc. | March, 2015 | Provide a Social Shopping Sites to Provide Personalized Shopping Experiences |
| Quickfire Networks | January, 2015 | Provide Video Format Conversion Services without Distortion |
| Wit.ai | January, 2015 | Provide for Turning Speech into Actionable Data |

| Google (2015-2019) | | |
|---|---|---|
| **Key Businesses** | **Time** | **Key Services** |
| Typhoon Studios | December, 2019 | Independent Game Development and Sales |
| CloudSimple | November, 2019 | Use Vmware to Compute Workload Based on Server Virtualization Technology |
| Socratic | October, 2019 | Offer Community that Provides Learning Opportunities for Students |
| Elastifile | July, 2019 | Develop File Storage Solutions for Google Cloud, Amazon Web Services, and Microsoft Azure |
| Looker | June, 2019 | Provides Data Analysis Services, Including Data Capture and Visualization |
| Alooma | February, 2019 | Move Company Data From Multiple Sources to a Single Database |
| Superpod | January, 2019 | Provide Professional Answers to Users' Questions |
| Sigmoid Labs | December, 2018 | Offers Train Running Times and Ticketing Purchases |
| Workbench | November, 2018 | Provide a Library of Online Courses Organized by Grade and Subject |
| Onward | October, 2018 | Use Natural Language to Extract Customer Content and Provide Personalized Answer Services |
| Senosis | September, 2018 | Use phones' Camera to Collect Health Data and Diagnose Health Conditions |
| GraphicsFuzz | August, 2018 | Develop Mobile Graphics Run Test Tools |
| Cask | May, 2018 | Provide Solutions for Large Data Analysis Services Based on Hadoop |
| Velostrata | May, 2018 | Provide Services to Migrate Local Data to the Cloud |
| Tenor | March, 2018 | Provide GIF Search Services |
| Redux | January, 2018 | Develop Technology that Uses Vibrations to Turn the Surface of Phone into a Speaker or to Provide Tactile Feedback |
| 60db | October, 2017 | Provide Personalized Short Video Content |
| Relay Media | October, 2017 | Provide Services to Convert Regular Web Pages to AMP Format |
| Bitium | September, 2017 | Provides Enterprise-class Identity Management and Access Tools for Cloud-based Applications |
| HTC | September, 2017 | Develop Smart Phones and VR Services |
| AIMatter | August, 2017 | Application that Uses Artificial Intelligence to Modify Users' Selfie |
| Halli Labs | July, 2017 | Develop Deep Learning and Machine Learning Systems |
| Owlchemy Labs | May, 2017 | Create Independent Cellular Networks for Rural Communities. |
| AppBridge | March, 2017 | Provide Real-time 3D Scene Reproduction Services |
| Kaggle | March, 2017 | Social Shopping Sites to Provide Personalized Shopping Experiences |
| Fabric | January, 2017 | Provide Video Format Conversion Services without Distortion |
| Limes Audio | January, 2017 | Develop Virtual Reality Games |
| Cronologics | December, 2016 | Develop Smartwatch Operation Systems |
| Qwiklabs | November, 2016 | Provide Practical Teaching Services for Cloud Platform and Basic Software Providers |
| LeapDroid | November, 2016 | Develop PC Version of Android Emulator |
| Eyefluence | October, 2016 | Provides Services to Extract Data Points from Eye Photos to Identify People |
| FameBit | October, 2016 | Establish Relationships with Merchants to Gain Access to Youtube's Internet Stars |
| API.AI | September, 2016 | API to Help Developers Create Conversational Chatbots |
| Urban Engines | September, 2016 | Provide Spatial Analysis Services to Help Local Governments and Enterprises Understand Information Related to Urban Development and Change, and to Help Improve the Level of Transportation |
| | | Services in Urban Areas |
| Apigee | September, 2016 | Provide API Products and Technologies for Enterprises and Developers |
| Orbitera | August, 2016 | Provide Channel Simplification for Cloud Business |
| LaunchKit | July, 2016 | Provides Developers with Application Design, Operation, and Management Services |
| Kifi | July, 2016 | Collect and Search Links Shared in Social Applications and Provide more Link Recommendations for Users |
| Anvato | July, 2016 | Provides Video Coding, Editing, Publishing, and Cross-platform Distribution Services |
| Moodstocks | July, 2016 | Develop Machine Learning-based Image Recognition Technology for Smartphones |
| Webpass | June, 2016 | Develop High Speed and Extensive Internet Connection for Residential and Commercial Buildings |
| Synergyse | May, 2016 | Provide Companies with Tutorials on How to Use Google Products |
| Pie | February, 2016 | Provide Communication Services for Teams |
| BandPage | February, 2016 | Offer Music Artists Chance to sell vips, party tickets et al |
| bebop | November, 2015 | Develop and Maintain Application Services for Developers |
| Fly Labs | November, 2015 | Provide Users with Photos and Videos Editing Services |
| Digisfera | October, 2015 | Provide 360-degree Image Photography, Design and Development Services |
| Agawi | June, 2015 | Provide Different System Application Transfer Services |
| Jibe Mobile | September, 2015 | Provide Cloud Platform that Connects Two Applications and Allows Real-time Data Transfer |
| Oyster | September, 2015 | Offer Ebook Subscriptions and Retail Services |
| Pixate | July, 2015 | Develop Tools for Developers to Prototype Their Mobile |
| | | Applications |
| Pulse.io | May, 2015 | Provides a Mobile App Performance Service that Helps Developers Identify Issues Slowing Down Production Apps. |
| Timeful | May, 2015 | Provide Time Management Services |
| Skillman & Hackett | April, 2015 | Provide Rapid Prototyping and Virtual Reality Services |
| Thrive Audio | April, 2015 | Develop 3D Virtual Audio Technology |
| Red Hot Labs | February, 2015 | Create Tools for App Developers that People Wished We Had. |
| Softcard | February, 2015 | Provide Mobile Payment Platform |
| Odysee | February, 2015 | Provide Picture and Video Backup, Synchronization, Sharing Services |
| Launchpad Toys | February, 2015 | Provide a Platform for Children to Create and Share Stories |

| **BIDU (2015-2019)** | | |
|---|---|---|
| **Key Businesses** | **Time** | **Key Services** |
| Kangfuzi | February, 2019 | Study the AI Knowledge Graph to Assist Doctors in Clinical Decision-making |
| KITT.AI | July, 2017 | Develop Speech Arousal and Natural Speech Interaction Technology |
| XPerception | April, 2017 | Provide Visual Perception Software and Hardware Solutions for applications |
| RavenTech | February, 2017 | Provide Smart Home Hardware Business |
| Li Jiaoshou | December, 2016 | Official Account Marketing |
| popln | June, 2015 | Support Native Advertising for News Media |
| Anquanbao | April, 2015 | Provide Users with Enterprise-class Application Security Firewall, Provide Acceleration and Security Services for Developers |

| Paypal (2015-2019) | | |
|---|---|---|
| **Key Businesses** | **Time** | **Key Services** |
| Honey | November, 2019 | Build Browser Plugins for Customers |
| GoPay | September, 2019 | Provide Mobile Payment Services |
| Simility | June, 2018 | Collect and Analyze Data for Fraud Officers |
| Hyperwallet | June, 2018 | Develop Online Employee Payment Platform Solutions for Growing Organizations and Marketing Companies Worldwide |
| iZettle | May, 2018 | Provide mobile payment services |
| Swift Financial | August, 2017 | Provide Working Capital for Small Business |
| TIO Networks | February, 2017 | Provide Multi-channel Payment Processing and Accounts Receivable Management Services |
| Modest Inc | August, 2015 | Provide Mobile Payment Services |
| Xoom Corporation | July, 2015 | Provide International Money Transfer Services |
| CyActive | March, 2015 | Provide Early Warning and Destroy Virus Services |
| Paydiant | March, 2015 | Provides Mobile Payment Services for Large Retail Chains |

| eBay (2015-2019) | | |
|---|---|---|
| **Key Businesses** | **Time** | **Key Services** |
| Motors.co.uk | October, 2018 | Provide Used Car Classified Advertising Information for Customers |
| Terapeak | December, 2017 | Selection, Sales Timing and Pricing Strategy Services for eBay Sellers |
| Corrigon | October, 2016 | Develop Computer Vision and Visual Search Techniques |
| SalesPredict | July, 2016 | Forecast Ideal Potential and Development Prospects of the Business for the Enterprise |
| Ticketbis | May, 2016 | Provide Online Ticketing Services |
| ExpertMaker | May, 2016 | Develop Professional Search Tools |
| Cargigi | March, 2016 | Provides Online Advertising and Marketing Services for the Automotive Industry |
| Twice | July, 2015 | Provide Second-hand Clothes Sales Platform |
| Vivanuncios | January, 2015 | Provide Second-hand Trading Platform |

**JD (2015-2019)**

| **Key Businesses** | **Time** | **Key Services** |
|---|---|---|
| Doraemon | September, 2019 | Provide Smart Payment and Marketing Solutions for Offline Merchants |
| Beijing Jade Palace Hotel | February, 2019 | Provide Guest Room, Catering, Entertainment, Shopping, Office Services |
| Farftech | June, 2017 | Privide Online Luxury Fashion Retail Platform |
| QG E-sports Club | June, 2017 | Participate in Various E-sports Competitions |
| TQMALL | March, 2017 | Provide auto parts procurement platform |
| YHD.com | June, 2016 | Provide online supermarket platform |
| Thumb Reading | June, 2015 | Reading platform for high-end people |

**Twitter (2015-2019)**

| **Key Businesses** | **Time** | **Key Services** |
|---|---|---|
| Aiden.ai | November, 2019 | Provide an AI Analytics Platform Intended to Assist Marketers How to Spend Their Budget Efficiently |
| Fabula AI | June, 2019 | Detect the Spread of False Information Online |
| Highly | April, 2019 | Provide Article Essence Sharing Platform |
| Smyte | June, 2018 | Provides Tools to Block Online Abuse, Harassment, Spam, and Protect User Accounts |
| Yes, Inc | December, 2016 | Develop Apps that Would Connect People Both in Real Life and from Afar |
| Magic Pony Technology | June, 2016 | Provide Expanded Data for Images Used |
| Peer | April, 2016 | Creates a Safe, Trusted Way for People to Share Anonymous Feedback. |
| ZeroPush | October, 2015 | Offer Developers to Integrate Push Notifications into Their Apps |
| Fastlane | October, 2015 | Provide Continuous Delivery Solutions for Ios and Android Applications |
| Whetlab | June, 2015 | Develop AI to Make Learning Faster for Artificial Intelligence |
| TellApart | April, 2015 | Offer Retailers to Leverage Their Data to Personalize the Customer Experience and Drive Omnichannel Commerce |
| TenXer | April, 2015 | Provide a Better Platform for Developers and Engineers to Collaborate |
| Periscope | March, 2015 | Provide Live Streaming Media Platform |
| Niche | February, 2015 | Provide Online Community that Enables Social Media Creators to Engage, Thrive, and Monetize |
| ZipDial | January, 2015 | Offers a Marketing and Analytics Platform that Enables Brand Managers to Drive Customer Engagement by Capitalizing on Missed Calls |

| Tencent (2015-2019) | | |
|---|---|---|
| **Key Businesses** | **Time** | **Key Services** |
| Sharkmob | May, 2019 | Develop Multiplayer Games with Social Elements and Competitiveness |
| Amer Sports | March, 2019 | Sports Goods Production |
| Fatshark | January, 2019 | Independently Develop Games |
| Cat Cake | August, 2018 | Provide Short Video Creation and Sharing Platform |
| Grinding Gear Games | May, 2018 | Develop Online Action Role-Playing Games |
| LCS | March, 2018 | Provide Overall Information Security Solutions |
| Sanook | December, 2016 | Provide Consumers with Entertainment, Information, Business and Community Services |
| Kuwo | July, 2016 | Provide Music Playing Platform |
| Kugou | July, 2016 | Provide Solutions for Internet Users and Digital Music Industry Development |
| Supercell | June, 2016 | Develop and Release Mobile Games |
| Riot Games | December, 2015 | Develop and Release Games |
| Miniclip SA | February, 2015 | Develop and Release Games |

| YRD (2015-2019) | | |
|---|---|---|
| **Key Businesses** | **Time** | **Key Services** |
| Daokoudai | July, 2019 | Provide P2P Lending Services Featuring Alumni Relationships |
| Hui Min | March, 2019 | Provide Unsecured Consumer Loans, Secured Consumer Loans, Financial Lease Transactions and Loans to Smes |
| Wealth Management | October 2016 | Provide Investment Product Advisory Services |
| Heng chen | February 2015 | Provide Unsecured Consumer Loans |

| **PPDF (2015-2019)** | | |
|---|---|---|
| **Key Businesses** | **Time** | **Key Services** |
| Ledao | November, 2018 | Provide Loan Services |
| Nianqiao | November, 2018 | Sports Goods Production |
| Zihe | July, 2017 | Provide Technology Development, Technology Consulting, Technology Services, Technology Transfer |
| Guangjian | June, 2017 | Provide Loan Services |

| **DNJR (2015-2019)** | | |
|---|---|---|
| **Key Businesses** | **Time** | **Key Services** |
| Xingjiuhao | October, 2018 | Produce and Sell for Internet of Things Technology and Technical Consulting |
| Youwang | April 2018 | Provide car rental business |
| Baoxun | February 2017 | Provide Online Marketing Design and Production of Online Advertising, Market Research Services |